\documentclass[useAMS,usenatbib]{mn2e}
\usepackage{fmtcount}
\usepackage{threeparttable}
\usepackage{graphicx,subfigure}
\usepackage{url}
\usepackage{rotating}

\title[Velocities of Type II-P SNe]{Measuring Expansion Velocities in Type II-P Supernovae}
\author[Tak\'ats \& Vink\'o]{K. Tak\' ats$^{1}$\thanks{E-mail: ktakats@titan.physx.u-szeged.hu (KT);\newline vinko@titan.physx.u-szeged.hu (JV)} 
and J. Vink\' o$^{1,2}$\footnotemark[1]\\
$^{1}$Department of Optics $\&$ Quantum Electronics, University of Szeged, D\' om t\' er 9, Szeged, Hungary\\
$^{2}$Department of Astronomy, University of Texas at Austin, Austin, TX 78712, USA\\
}

\begin{document}

\date{Accepted  Received ; in original form }

\pagerange{\pageref{firstpage}--\pageref{lastpage}} \pubyear{}

\maketitle

\label{firstpage}

\begin{abstract}

We estimate photospheric velocities of Type II-P supernovae using model 
spectra created with {\tt SYNOW}, and compare the results with those obtained by more conventional techniques, such as cross-correlation,
or measuring the absorption minimum of P Cygni features. 
Based on a sample of 81 observed spectra of 5 SNe, we show that {\tt SYNOW} provides velocities that are similar to ones
obtained by more sophisticated NLTE modeling codes, but they can be derived in a less computation-intensive way.
The estimated photopheric velocities ($v_{model}$) are compared to ones measured from Doppler-shifts of the absorption minima of the H$\beta$ and the Fe\,{\sc ii} $\lambda5169$ features. 

Our results confirm that the Fe\,{\sc ii} velocities ($v_{Fe}$) have tighter and more homogeneous correlation with the  estimated photospheric velocities than the ones measured from H$\beta$, but both suffer from phase-dependent systematic deviations from those. The same is true for comparison with the cross-correlation velocities. 
We verify and improve the relations between $v_{Fe}$, $v_{H\beta}$ and $v_{model}$ in order to provide useful
formulae for interpolating/extrapolating the velocity curves of Type II-P SNe to phases not covered by observations.   
We also discuss the implications of our results for the distance measurements of Type II-P SNe, and
show that the application of the model velocities is preferred in the Expanding Photosphere Method.

\end{abstract}


\section{Introduction}\label{sec_intro}

Type II-P supernovae (SNe) are core-collapse (CC) induced explosions that eject massive hydrogen-rich
envelope. These SNe show a nearly constant-luminosity plateau in their light curve.
The plateau phase lasts for $\sim$100 days, and finally ends up in a rapid decline
of the luminosity as the expanding ejecta becomes fully transparent.
  
CC SNe and their progenitors play important role in 
understanding the evolution of the most massive stars. Extensive searches for 
SN progenitors showed that the progenitors of Type II-P SNe are stars with 
initial masses between $\sim$ 8 and 17 M$_\odot$ \citep{smarttprog}. 
The lower value is roughly in agreement with the prediction of current theoretical evolutionary
tracks, but the upper limit is somewhat lower than expected. The fate of the more massive
(M $\geq$ 20 M$_\odot$) stars is controversial. They are thought to be the input channel
for Type Ib/c SNe \citep{gaskell86}, but this has not been confirmed directly by
observations yet \citep{smartt_rev}. 

Any observational study of the physics of SNe and their progenitors (and, obviously, all astronomical objects) 
depends heavily on the knowledge of their distances. Because distance is such a crucial parameter, 
strong efforts have been devoted to the development of distance measurement techniques that are applicable
for Type II-P SNe.  
 
The Expanding Photosphere Method (EPM) \citep{kirshner_epm}, a variant of the Baade-Wesselink method, 
uses the idea of comparing the angular and the physical diameter of the expanding SN ejecta. The input quantities 
are fluxes, temperatures and velocities observed at several epochs during the photospheric phase.
Thus, EPM requires both photometric and spectroscopic monitoring of SNe throughout the plateau phase. 
A clear advantage of EPM is that it does not require external calibration via SNe with known distances.
Another interesting property is that EPM is much less sensitive to uncertainties in the interstellar 
reddening and absorption toward SNe. As \citet*{E96} (hereafter E96) have shown, 1 mag uncertainty
in $A_V$ results in only $\sim 8$ \% error in the derived distance. However, the assumption that the
SN atmosphere radiates as a blackbody diluted by electron scattering raised some concerns. These led to the
development of the applications of full NLTE model atmospheres, like the PHOENIX code in the 
``Spectral-fitting Expanding Atmosphere Method'' (SEAM) \citep{baron99em}, 
or the CMFGEN code by \citet*{dessart2005a, dessart2005b}. The drawback of these approaches is that
the building of tailored model atmospheres can be very time-consuming and needs much more computing power. Also, in order to compute reliable models, the input observed spectra need to have sufficiently high S/N and spectral resolution.

The more recently developed Standardized Candle Method (SCM) \citep{hamuy_pinto_scm, hamuy_scm} 
relies on the empirical correlation between the measured expansion velocity and the 
luminosity in optical \citep{nugent06, poznanski09} or in NIR \citep{maguirescm} bands in
the middle of plateau phase, at +50 days after explosion. This method requires less extensive input 
data, but needs a larger sample of Type II SNe with 
independently known distances in order to calibrate the empirical correlation. Because this is basically
a photometric method which compares apparent and absolute magnitudes, SCM is more sensitive to 
interstellar reddening than EPM/SEAM, as mentioned above.

Beside photometry, both EPM/SEAM and SCM need information on the expansion velocity at the 
photosphere ($v_{phot}$). At first, EPM seems to be more challenging, because it requires multi-epoch
observations, while SCM needs only the velocity at a certain epoch, at +50 days after explosion
($v_{50}$). However, direct measurement of $v_{50}$ is possible only in the case of
precise timing of the spectroscopic observation, which is rarely achievable. 

The correct determination of $v_{phot}$ is not trivial.    
As SNe expand homologously ($v \sim r$, where $v$ is the velocity of a given layer and $r$ is 
its distance from the center), in most cases it
is difficult to derive a unique velocity from the observed spectral features. 
The most frequently followed approach relies on measuring the Doppler-shift of
the absorption minimum of certain spectral features (mostly Fe\,{\sc ii} $\lambda 5169$ or H$\beta$).
Another possibility is the computation of the cross-correlation between the SN spectrum
and a set of spectral templates, consisting of either observed or model spectra, with
known velocities. The third method is building a full NLTE model (like PHOENIX or CMFGEN) 
for a given SN spectrum, and adopting the theoretical $v_{phot}$ from the best-fitting model.

The aim of this paper is to present a similar, but less computation-demanding approach 
to assign velocities to observed SN spectra. We apply the simple parametrized code {\tt SYNOW} 
\citep{fisher99, hatano99} to model the observed spectra with an
approximative, but self-consistent treatment of the formation of lines
in the extended, homologously expanding SN atmosphere. To illustrate the applicability of
{\tt SYNOW}, we construct and fit parametrized models to a sample of five well-observed, nearby Type II-P
SNe that have a series of high S/N spectra publicly available. We also check and re-calibrate
some empirical correlations between velocities derived from various methods.
The description of the observational sample is given in \S \ref{sec_data}. In \S \ref{sec_issue} we first 
review the different velocity measurement techniques for SNe II-P, then we present the details of the 
application of {\tt SYNOW} models (\S \ref{sec_synow}). The results are collected and 
discussed in \S \ref{sec_results} and \S \ref{sec_disc}, while the implications of the results 
for the distance measurements are in \S \ref{sec_dist}. We draw our conclusions in \S \ref{sec_concl}.

\section{Data}\label{sec_data}

Our sample contains 81 plateau phase--spectra of five objects, SNe 1999em, 2004dj, 2004et, 2005cs and 2006bp, respectively. All five SNe are well-observed objects, they have been studied in detail, and show a wide variety in their physical properties (see Table \ref{physdata}). 

\begin{table*}
 \centering
 \begin{minipage}{145mm}
  \caption{The physical properties of the supernovae used in this paper.}
  \label{physdata}
  \begin{tabular}{@{}lccccccc@{}}
  \hline
   SN & t$_0$ & Distance & cz\footnotemark[1] & E(B-V)   & $M_{Ni}$  & $M_{prog}$ & References\footnotemark[2]\\
       & (JD-245000) &  (Mpc) & (km s$^{-1}$) & (mag) & ($10^{-2}$ M$_\odot$) & (M$_\odot$) & \\
 \hline
 SN 1999em & 1477.0 & 7.5 -- 12.5 & 717 & 0.10  & 2.2 -- 3.6 & $<$15  & 1,2,3,4,5,6,7,8 \\
 SN 2004dj & 3187.0 & 3.2 -- 3.6 & 131 & 0.07  & 1.3 -- 2.2 & 12 -- 20 & 9,10,11,12,13,14,15\\
 SN 2004et & 3270.5 & 4.7 -- 6.0 & 48 & 0.41  & 5.6 -- 6.8 & 9, 15 -- 20 & 16,17,18,19,20,21,22 \\
 SN 2005cs & 3549.0 & 7.1 -- 8.9 & 463 & 0.05  & 0.3 -- 0.8 & 6 -- 13 & 23,24,25,26,27,28,29\\
 SN 2006bp & 3835.1 & 17.0 -- 18.3 & 987 & 0.40 & -- & 12 -- 15 & 26,30 \\
\hline
\end{tabular}
 \begin{tablenotes}
       \item[a]{$^1$ NED, \url{http://nedwww.ipac.caltech.edu/}}
       \item[b]{$^2$ References: (1) \citet{hamuy99em}, (2) \citet{leonard99em}, (3) \citet{smartt99em}, (4) \citet{leonard_ceph}, (5) \citet{elmhamdi99em}, (6) \citet{baron99em}, (7) \citet{dessart2006}, (8) \citet{utrobin99em}, (9) \citet{MA04dj}, (10) \citet{kotak04dj}, (11)  \citet{wang04dj}, (12) \citet{chugai04dj}, (13) \citet{zhang04dj}, (14) \citet{vinko04dj}, (15) \citet{vinko04dj2}, (16) \citet{li04et}, (17) \citet{sahu04et}, (18) \citet{misra04et}, (19) \citet{utrobin04et}, (20) \citet{poznanski09}, (21) \citet{maguire04et}, (22) \citet{crockett04et}, (23) \citet*{maund05cs}, (24) \citet{pastorello05csI}, (25) \citet{takats05cs}, (26)  \citet{dessart2008}, (27) \citet*{eldridge05cs}, (28) \citet{utrobin05cs}, (29) \citet{pastorello05csII}, (30) \citet{immler06bp}}
     \end{tablenotes}
\end{minipage}
\end{table*}

SN 1999em was discovered on 29th October 1999 by \citet{discov99em} in NGC 1637 at a very early phase. 
It is a very well-observed, well-studied object The explosion date was determined as $2451477.0 \pm 2$ JD 
\citep{leonard99em, hamuy99em, dessart2006}. 
We used the spectra of \citet{leonard99em} and \citet{hamuy99em} downloaded from the 
{\tt SUSPECT}\footnote{http://suspect.nhn.ou.edu/$\sim$suspect/} database. 
Our sample contains 22 spectra, covering the first 80 days of the plateu phase.

SN 2004dj was discovered on 31st July 2004 by Itagaki \citep{discov04dj} in a young, massive cluster Sandage-96 of NGC 2403, about 1 month after explosion. We used the spectra taken by \citet{vinko04dj}. Due to the lack of observed spectrophotometric standards, the flux-calibration of those spectra was inferior, but this is not a major concern when only velocities are to be determined.
We included 12 spectra taken between +47 and +100 days after explosion in our sample.

SN 2004et was discovered by Moretti \citep*{discov04et} on 27th September 2004 in NGC 6946. 
The spectra of \citet{sahu04et} (downloaded from {\tt SUSPECT}) and \citet{maguire04et} were used, 
together with 6 previously unpublished early-phase spectra taken with the 1.88-m telescope at DDO 
(see the Appendix \ref{appendix_a}).
The 22 spectra cover the period of $+$11 to $+$104 days after explosion.

SN 2005cs was discovered on 29th June 2005 by \citet{discov05cs} in M51. Due to its very early discovery and proximity, this object is very well-sampled and studied in detail. It is a underluminous, low-energy, Ni-poor SN that had a low-mass progenitor (see Table \ref{physdata} for references).
14 spectra of \citet{pastorello05csI} and \citet{pastorello05csII} were used, which were obtained between days +3 and +61.

SN 2006bp was discovered on 9th April 2006 by \citep{discov06bp} in NGC 3953, also in a very early phase. 
We used 11 spectra of \citet{quimby06bp}, downloaded from {\tt SUSPECT}, covering the period between $+$5 and $+$72 days.

\section{The issue of measuring expansion velocities in SNe}\label{sec_issue}

During the photospheric phase the homologously expanding Type II-P SN ejecta consist of two parts: the outer, partly transparent atmosphere where the observable spectral features are formed, and an optically thick inner part, which emits most of the continuum radiation as a ``diluted'' blackbody (e.g. \citeauthor{E96}). Because the inner part is mostly ionized, the major source of the total opacity is electron scattering. Thus, the {\it instantaneous} photosphere is located at the depth where the outgoing photons are last scattered, i.e. where the electron-scattering optical depth is $\tau_e \sim 2/3$ \citep*[][hereafter D05]{dessart2005a}. This is presumed to occur in a thin, spherical shell at a certain radius $r_{phot}$, and the  homologous expansion gives this a unique velocity $v_{phot} = v_{ref} \cdot r_{phot} / r_{ref}$, where $r_{ref}$ is the radius of an (arbitrary) reference layer and $v_{ref}$ is its expansion velocity. As the ejecta expands and dilutes, the photosphere migrates inward, toward the inner layers of the ejecta that expand slower, thus, $v_{phot}$ continuously declines in time.

The issue of measuring $v_{phot}$ comes from the fact that no measurable spectral feature is connected directly to $v_{phot}$. As previously mentioned in \S \ref{sec_intro}, there exist several observational/theoretical approaches to {\it estimate} $v_{phot}$ from observed spectra. In the followings we briefly review these methods, summarizing their advantages/drawbacks.

\subsection{P Cygni lines}\label{sec_pcyg}   

The most widely used method is to measure the velocity represented by the minimum flux of the absorption component of an unblended P Cygni line profile \citep[sometimes referred to as ``maximum absorption'', $v_{abs}$,][]{dessart2005b}. This is motivated by the theory of P-Cygni line formation \citep[e.g.][]{kasen02}, which predicts that the minimum flux in a P-Cygni line occurs at the velocity of the photosphere. Strictly speaking, this is only true for an optically thin line formed by pure scattering. In most cases the Fe\,{\sc ii} $\lambda 5169$ feature is used for measuring $v_{abs}$, for which \citet{dessart2005b} pointed out that it can represent the true $v_{phot}$ within 5 -- 10 \% accuracy. However, \citet{leonard99em} showed that using weaker, unblended features one can get $\sim 10$ \% lower velocities than from the Fe\,{\sc ii} $\lambda 5169$  feature. This raises the question of whether the measurable lines were indeed optically thin and unblended, and which one represented the true $v_{phot}$. Moreover, \citet{dessart2005b} also revealed that $v_{abs}$ can either over- or underestimate $v_{phot}$, depending on various physical conditions, like the density gradient and the excitation/ionization conditions for the particular transition within the ejecta. 

While the Fe\,{\sc ii} $\lambda 5169$ feature can be relatively easily identified and measured in Type II-P SNe spectra obtained later than $\sim 20$ days after explosion (which is undoubtedly a major advantage of this method), this is not so for earlier phases. In early-phase spectra only the Balmer-lines and maybe He\,{\sc i} features (mostly the $\lambda 5876$ feature) can be identified. These features are less useful for measuring $v_{phot}$, because none of them are optically thin, and the He\,{\sc i} $\lambda 5876$ feature can sometimes be blended with Na\,{\sc i} D. Moreover, their line profile shapes often make the location of the minimum of absorption component difficult to determine \citep{dessart2005b, dessart2006}.  

Also, in spectra having less signal-to-noise, the weaker Fe\,{\sc ii} features are often buried in the noise, making them inappropriate for velocity measurement. In these cases several authors attempted to use the stronger features, especially $H\beta$. Although \citet{dessart2005b} pointed out that $v_{H\beta}$ is certainly less tightly  connected to $v_{phot}$ than the weaker Fe\,{\sc ii} $\lambda 5169$ feature,  \citet{nugent06} have shown that the ratio $v_{H\beta} / v_{Fe}$ is $\sim 1.4$ below $v_{H\beta} = 6000$ km s$^{-1}$, and it is linearly decreasing for higher velocities. Recently this correlation was revised by \citet*{poznanski} by revealing an entirely linear relation between $v_{H\beta}$ and $v_{Fe}$, but restricting the analysis only for velocities measured between +5 and +40 days after explosion. The uncertainty of estimating $v_{Fe}$ from $v_{H\beta}$ is $\sim 300$ km s$^{-1}$ \citep{poznanski}. It is emphasized that these empirical relations are more-or-less valid between the  velocities of two observable features, but both of them may be systematically off from the true $v_{phot}$. Indeed, \citet{dessart2005b} found that $v_{H\beta}$ can be either higher or lower than $v_{phot}$ with an overall scattering of $\pm 15$ \%. For velocities below 10,000 km~s$^{-1}$ $v_{H\beta}$ is usually higher than $v_{phot}$, in accord with the empirical correlations with $v_{Fe}$, but above 10,000 km~s$^{-1}$ it can be lower than $v_{phot}$.     
 
\subsection{Cross-correlation}\label{sec_cross}

\begin{figure*}
\begin{center}
\includegraphics[]{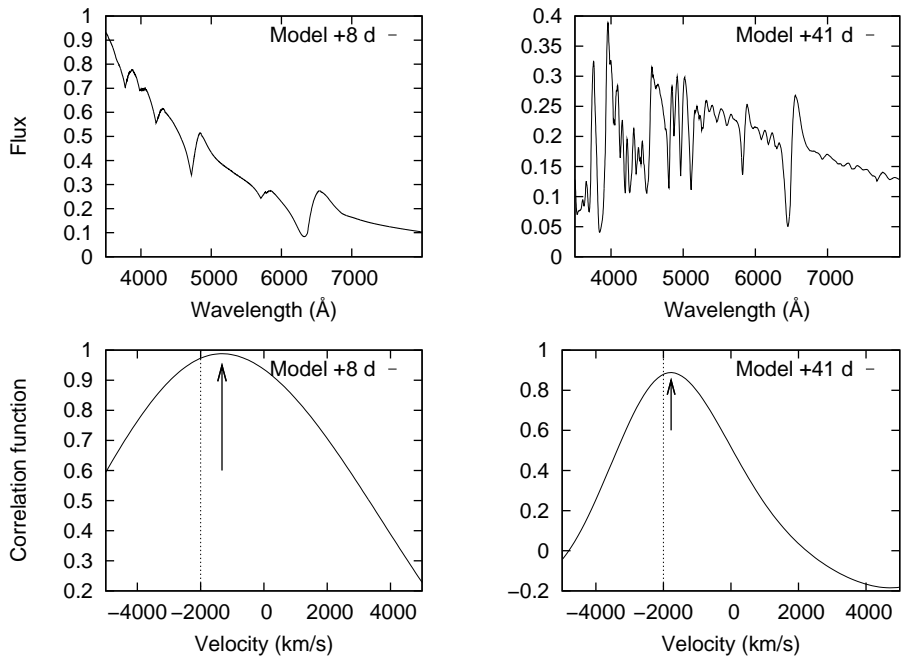}
\caption{Model spectra of a Type II-P SN in early-phase (top left) and late-phase (top right), and the 
resulting CCFs (bottom panels) after cross-correlating the spectra with their identical models but 
having 2000 km~s$^{-1}$ higher $v_{phot}$. The arrows mark the peak location of the CCFs, while
the true relative velocity difference $\Delta v_{phot}$ is indicated by the dotted vertical line. The peak of the
CCF underestimates $\Delta v_{phot}$ in both cases.}
\label{ccf}
\end{center}
\end{figure*}

Motivated by the uncertainties of measuring the Doppler-shift of a single, weak feature in a noisy spectrum, several authors proposed the usage of the cross-correlation technique, which is a powerful method for measuring Doppler-shifts of stellar spectra containing many narrow spectral features. This method predicts the Doppler-shift of the {\it entire} spectrum by computing the cross-correlation function (CCF) between the observed spectrum and a template spectrum with {\it a-priori} known velocity $v_{templ}$. The resulting velocity is $v_{cc} = v_{rel} + v_{templ}$, where $v_{rel}$ is the velocity where the CCF reaches its maximum (i.e. the relative Doppler-shift between the observed spectrum and the template).  

However, the applicability of cross-correlation for SN spectra is not obvious, because of the physics of P Cygni line formation in SN atmospheres. If $v_{phot}$ is higher, then the center of the absorption component gets blueshifted, while the center of the emission component stays close to zero velocity. Thus, cross-correlating two spectra with different $v_{phot}$, the relative velocity will underestimate the true velocity difference between the two spectra.  This is illustrated in Fig.~\ref{ccf}, where we determined the CCF between two Type II-P SN model spectra (computed with {\tt SYNOW}, see below) that were identical except their $v_{phot}$ which differed by $\Delta v_{phot} = 2000$ km s$^{-1}$. The cross-correlation was computed only in the 4500 - 5500 \AA~ interval to exclude the $H\alpha$ feature.   This was done for an early-phase spectrum containing only H and He features (Fig.~\ref{ccf} left panel) and a later-phase spectrum with more developed metallic lines (Fig.~\ref{ccf} right panel). It is visible in the bottom panels that the maximum of the CCF (i.e. $v_{rel}$) is shifted toward lower velocities with respect to the true $\Delta v_{phot}$ (indicated by a dotted vertical line in the CCF plot) for both spectra.  The systematic offset of $v_{rel}$ from $\Delta v_{phot}$ is $\sim 200$ km~s$^{-1}$ for the later-phase spectrum that contains many narrower absorption  features, and it reaches $\sim 700$ km s$^{-1}$ for the early-phase spectrum that are dominated by broad Balmer lines with stronger emission component. These simple tests are in very good agreement with the results of \citet{hamuy99em} and \citet{hamuythesis}, who applied the cross-correlation technique  using model spectra of \citeauthor{E96} with {\it a-priori} known $v_{phot}$  as templates. They pointed out that $v_{rel}$ usually underestimates $\Delta v_{phot}$  so that $\Delta v_{phot} / v_{rel} \sim 1.18$, with the scatter of $\sim 900$ km s$^{-1}$. 

Another variant of the CCF method was proposed by \citet{poznanski09}, and it was also applied by \citet{dandrea} and \citet{poznanski}. They took their template spectra from  the library of the SuperNova IDentification code  
{\tt SNID}\footnote{http://marwww.in2p3.fr/$\sim$blondin/software/snid/index.html} \citep{snid} that contains high S/N observed spectra of many SNe, but used only those templates that showed well-developed Fe\,{\sc ii} $\lambda 5169$ feature. The reason for their choice was that they wanted to estimate $v_{Fe}$ at +50 days after explosion ($v_{50}$), which is an input parameter in SCM. This template selection caused the well-known issue of template mismatch that further biases the cross-correlation results. As \citet{dandrea} concluded, this template mismatch can result in significant underestimate of $v_{Fe}$ by $\sim 1500$ km s$^{-1}$ for spectra obtained at $t < 20$ days, and is also present in later-phase specta, although being less pronounced, $\sim 300 - 400$ km~s$^{-1}$.  To overcome this problem, in a subsequent paper \citet{poznanski} suggested the application of their  $v_{Fe}$ vs. $v_{H\beta}$ relation to estimate $v_{Fe}$ by measuring $v_{H\beta}$  for these early-phase spectra and then propagate the resulting velocity to day +50.   Nevertheless, the two groups presented velocities that were different by $200$ to $1000$ km~s$^{-1}$ for the same set of SNe spectra from the SDSS-II survey. This underlines that although cross-correlation seems to be an easy and robust method that gives reasonable velocities even for low S/N spectra,  the results may be heavily biased, especially at early phases, when the SN spectra contain only a few, broad features.

\subsection{Tailored modeling}\label{sec_tailored}

Tailored modeling of the whole observed spectrum via NLTE models
were invoked e.g. by \citet{baron99em} using the code {\tt PHOENIX}, and
\citet{dessart2006} and \citet{dessart2008} with the code {\tt CMFGEN}.
Here $v_{phot}$ is determined implicitly by requiring
an overall fit of the entire observed spectrum by a synthetic spectrum 
from a full NLTE radiation-hydrodynamics model. In these models the
location of the photosphere is usually defined as the layer where
the electron scattering optical depth reaches unity or 2/3 (e.g. \citeauthor{dessart2005a}).  $v_{phot}$ is then simply determined by the radius $r_{phot}$ and the law of homologous expansion (see above). 
Although this is probably the best self-consistent method for obtaining SN velocities,
building full NLTE models for multiple epochs requires a lot of computing power.
Thus, its usage for a larger sample of SNe would be very time-consuming and impractical. 
Also, in some cases a good global fit to the whole spectrum may not be equally good
for individual spectral lines, where lots of physical details play important role in the
formation of the given features. This may lead to some uncertainties in the
velocities given by the models, which will be illustrated in \S~\ref{sec_results}.

\subsection{Using {\tt SYNOW}}\label{sec_synow}

The success of tailored spectrum models to estimate $v_{phot}$ suggests a similar, but much more simplified approach: the usage of {\it parametrized} spectrum models that do not contain the computation-intensive details of NLTE level populations or exact radiative transfer calculations, but preserve the basic physical assumptions of an expanding SN atmosphere, and able to reproduce the formation of P Cygni lines  in a simplified manner. Such models can be computed either with the {\tt SYNOW} code \citep{branchsynow},  or the more recently developed {\tt SYNAPPS} code \citep{synapps} that has a parameter-optimizer routine built-in.  In this paper we apply {\tt SYNOW} to calculate model spectra. Because {\tt SYNOW} does only spectrum synthesis and has no fitting capabilities, we used self-developed UNIX shell scripts to fine-tune the parameters until a satisfactory fit to the observed spectrum is achieved.  
  
The basic assumptions of {\tt SYNOW} are the followings: $i)$ the SN ejecta expand homologously; $ii)$ the photosphere radiates as a blackbody;  $iii)$ spectral lines are formed entirely above the photosphere; $iv)$ the line formation is due to pure resonant scattering. Level populations are treated in LTE, and the radiative transfer equation is solved in the Sobolev approximation \citep[see also in e.g.][]{kasen02}.

When running {\tt SYNOW}, several parameters must be set. These are the temperature of the blackbody radiation ($T_{bb}$) emitted by the photosphere, the expansion velocity at the photosphere ($v_{phot}$), the chemical composition of the ejecta and the optical depth of a reference line ($\tau_{ref}$) of each compound. For each atom/ion the optical depths for the rest of the lines are calculated assuming Boltzmann excitation governed by the excitation temperature $T_{exc}$. 
The location of the line-forming region in the atmosphere can be tuned for each compound by setting the velocities of the lower and upper boundary layers, $v_{min}$ and $v_{max}$. The optical depth as a function of velocity (i.e. radius) can be modeled either as a power-law, or an exponential function. We assumed power-law atmospheres, and adjusted the power-law exponent $n$ to reach optimal fitting.

After setting the inital values by hand, several models in a wide range of $v_{phot}$, $n$, and $\tau_{ref}$ were created 
for a pre-selected set of ions. In order to reduce the
number of free parameters, we initially set $T_{bb}$ (which has very little effect on the line shapes) to represent the
continuum of the fitted spectrum and kept it fixed during the optimization. Moreover, we applied a single power-law exponent $n$ for all atoms/ions. We also assumed that all spectral features are photospheric, thus, fixing $v_{min}$ well below the photosphere and $v_{max}$ at $\sim 40000$ km s$^{-1}$.  

The best-fitting model was then chosen via $\chi^2$-minimization, and  
the fitting process was iterated for a few times, each time resampling the parameter 
grid in the vicinity of the minimum of the $\chi^2$ function found in the previous iteration cycle. 
This way we determined the parameters and the chemical compositions that best describe the observed spectra. 
 
Then, to further refine the estimated photospheric velocity, we fine-tuned only $v_{phot}$ 
of the best-fitting model, and calculated the $\chi^2$ function only in the vicinity of certain lines instead of the whole spectrum. This may reduce the systematic under- or overestimate of
$v_{phot}$ produced by false positive fitting to the observed spectrum outside the range of the considered spectral features. 

Motivated by the results of \citet{dessart2005b} (see \S \ref{sec_pcyg}), we have chosen the Fe\,{\sc ii} $\lambda5169$ 
feature for this fine-tuning process. When this feature was not present
in the observed spectrum (i.e. the early-phase spectra, before $\sim$15 days) we used 
H$\beta$ (see \S \ref{sec_pcyg}) instead.
Hereafter we denote the $v_{phot}$ parameter of the best-fitting {\tt SYNOW} model as $v_{model}$. 
Errors of $v_{model}$ were estimated by choosing the 90 \% confidence interval around the minimum of the $\chi^2$ function. 

\begin{figure*}
\includegraphics[width=140mm]{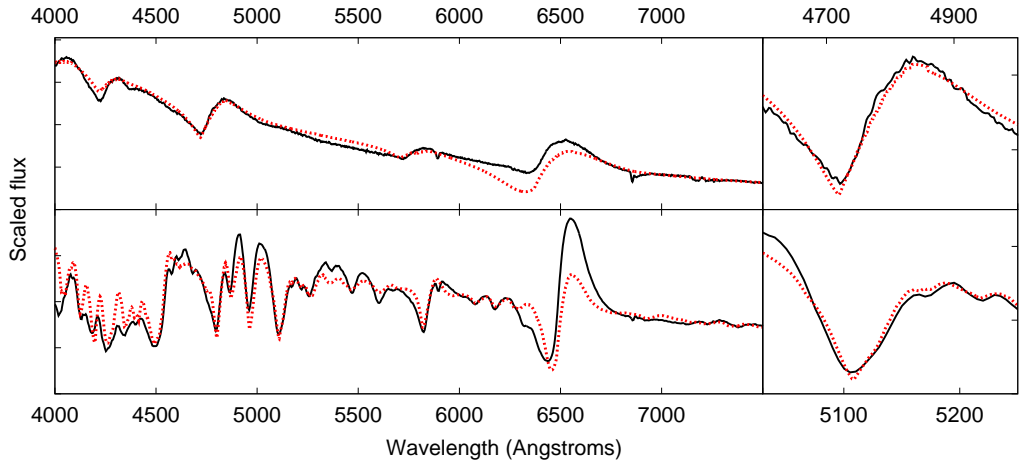}
\caption{The observed (black solid line) and best-fitting model (red dashed line) spectra of SN 1999em on days $+$9 (top) 
and $+$41 (bottom). The right panels enlarge the wavelength regions of the fitted lines: H$\beta$ (top) and Fe\,{\sc ii} $\lambda5169$ (bottom).}
\label{multisp}
\end{figure*}

Fig. \ref{multisp} shows two examples for an early- and a later-phase spectrum of SN 1999em  together with the best-fitting model. The right panel zooms in on the region of H$\beta$ and Fe\,{\sc ii} $\lambda5169$.  Note that although the final fitting was restricted to the proximity of these lines, the best model describes the entire observed spectrum (except H$\alpha$) very well.   

This velocity measurement method have multiple sources of error. One of them may be the systematic bias 
due the approximations in the model (LTE, power-law atmosphere, simple source function, etc.). However, the comparison of our results with those from full NLTE {\tt CMFGEN} models 
(\S \ref{sec_results}) show no systematic bias in the case of SNe 1999em and 2005cs. The agreement between the velocities 
from these two very different modeling codes are within $\pm 10$ percent.
For SN~2006bp the differences are higher, but it will be shown below that for this
SN the {\tt CMFGEN} models do not describe well the spectral features we use, contrary to 
the {\tt SYNOW} models (\S \ref{sec_results06bp}). 

\begin{figure*}
\begin{center}
\subfigure[]{
\label{contur_a}
\includegraphics[width=0.40\textwidth]{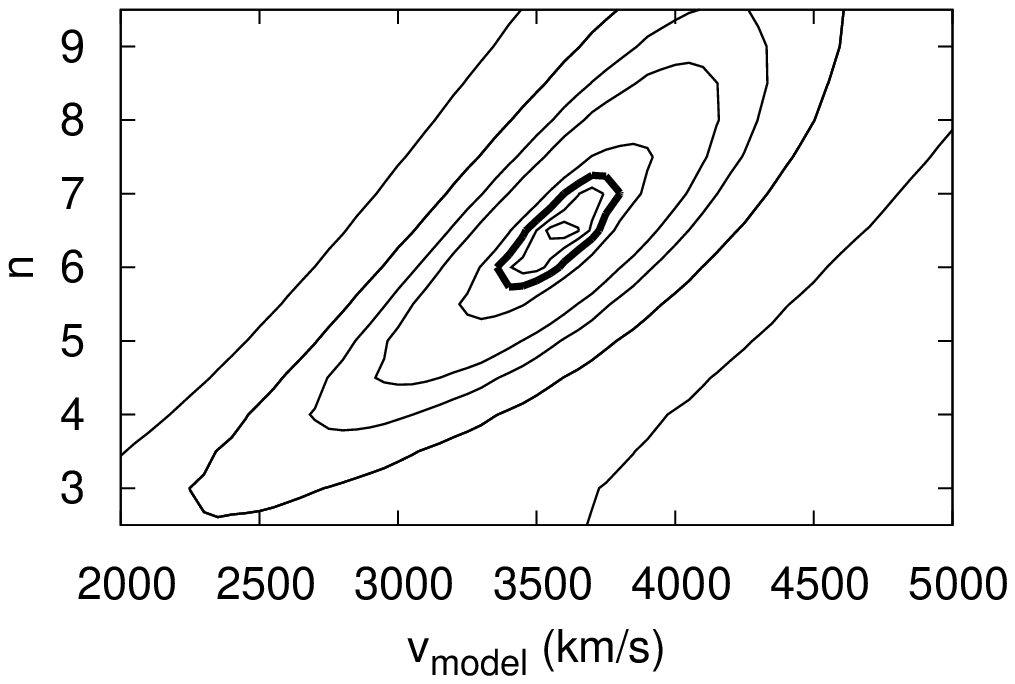} }
\subfigure[]{
\label{contur_b}
\includegraphics[width=0.40\textwidth]{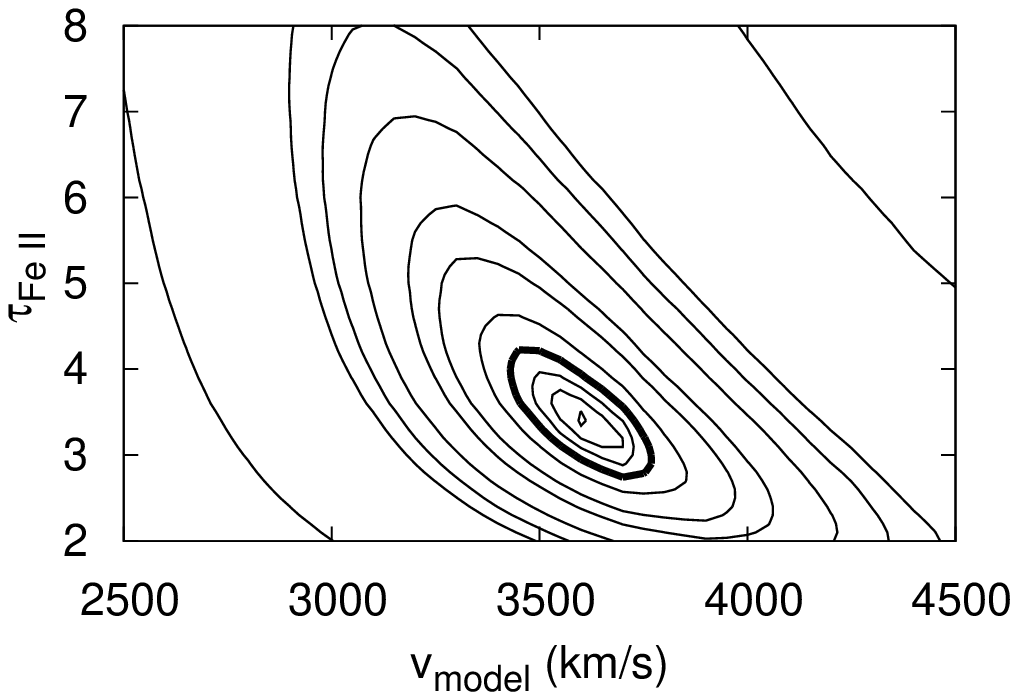}}\\
\subfigure[]{
\label{contur_c}
\includegraphics[width=0.40\textwidth]{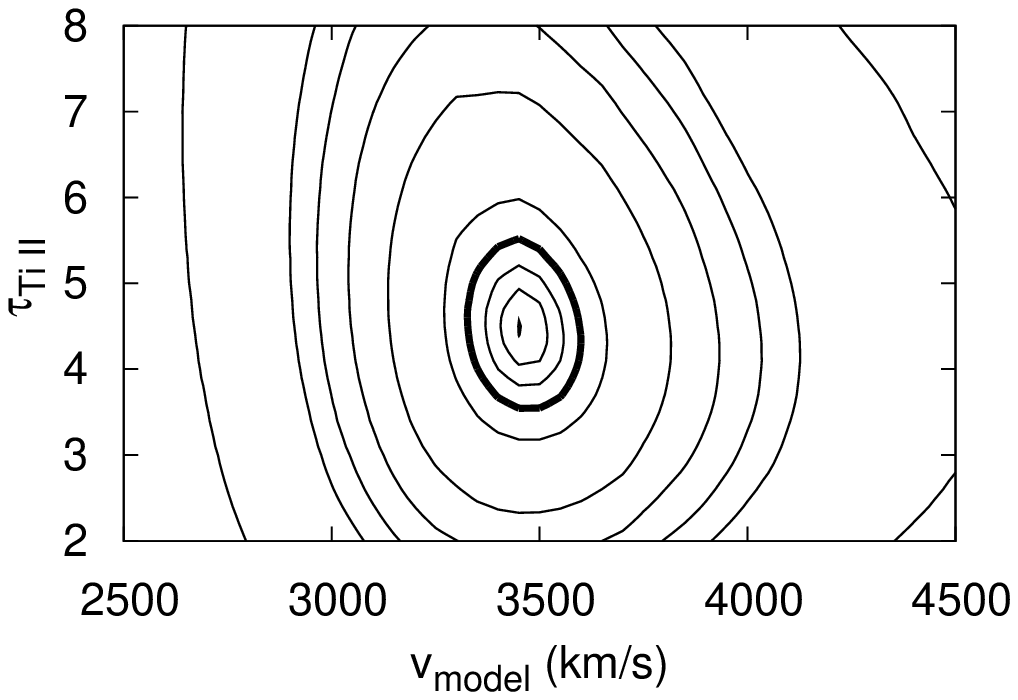}}
\subfigure[]{
\label{contur_d}
\includegraphics[width=0.40\textwidth]{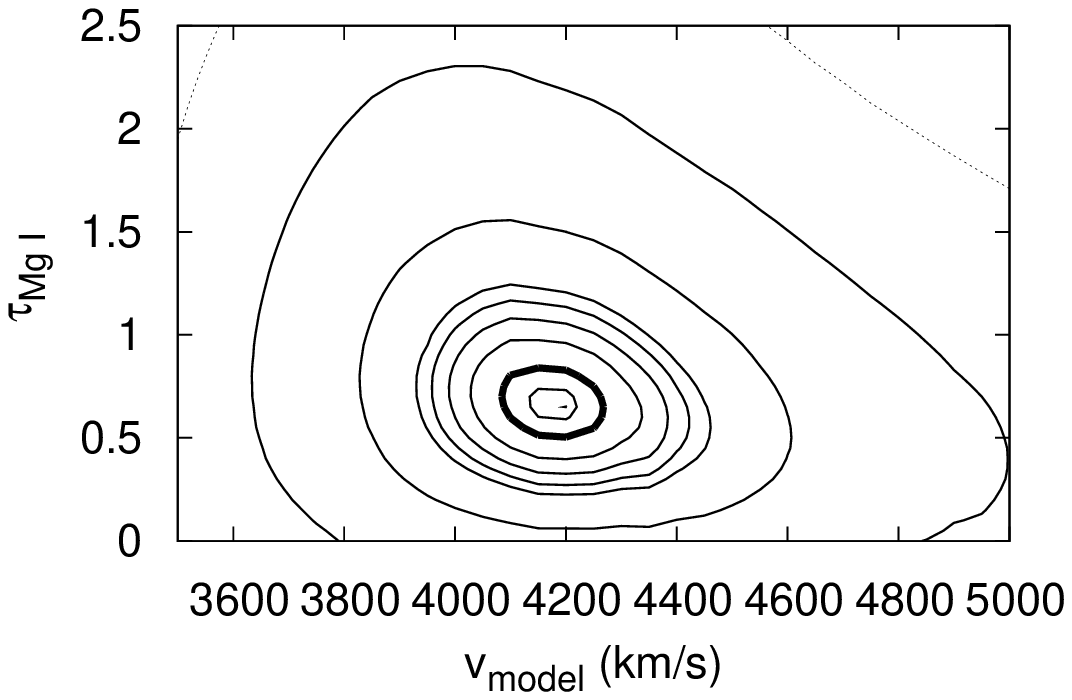}}\\
\caption{Contour plot of the $\chi^2$ function around its minimum, as a function of
$v_{model}$ and power-law exponent $n$ (a), $\tau_{ref}$ of Fe\,{\sc ii} (b), Ti\,{\sc ii} (c) and Mg\,{\sc i} (d). 
The thick black contour curve corresponds to 50 \% higher $\chi^2$ than the minimum value.
The parameters plotted in (a) and (b) are definitely correlated, while
the correlation is much less between the parameters in the (c) and (d) panels.}
\label{contur}
\end{center}
\end{figure*}

Another source of error may be the correlation between the parameters. 
In Fig.~\ref{contur} we present
contour plots of the $\chi^2$ hyperspace around its minimum, as a function of $v_{model}$ and
several other parameters that can affect the shape of the fitted Fe\,{\sc ii} $\lambda5169$ 
feature. The thick black contour curve corresponds to 50 \% higher $\chi^2$ than the minimum value.
It is visible that correlation is indeed present (i.e. the contours are distorted) 
between $v_{model}$ and the power-law exponent $n$ or the optical depth $\tau_{Fe}$. 
The correlation is much less  between $v_{model}$ and
$\tau_{ref}$ of Ti\,{\sc ii} and Mg\,{\sc i}, whose features may blend with
Fe\,{\sc ii} $\lambda5169$. However, even for the correlated parameters,
selecting $n$ or $\tau_{Fe}$ very far from their optimum value 
can alter $v_{model}$ only by a few hundred km s$^{-1}$.
Thus, we conclude that uncertainties in finding the minimum of $\chi^2$ do not 
cause errors in $v_{model}$ that significantly exceed the uncertainty due to the spectral resolution of 
the observed spectra (which is usually between 200 - 300 km s$^{-1}$).

A possible source of uncertainty may be that during the final fitting the wavelength interval around the used spectral feature is chosen somewhat subjectively. However, our tests showed that changing the limits reasonably has negligible effect on the final velocities. 

It is emphasized that although the final fitting is restricted to a vicinity of
a well-defined spectral line, this method is certainly more reliable than the
measurement of only the location of the {\it minimum} of the same feature. As it was
discussed above, the minimum can be significantly and systematically altered
by signal-to-noise, spectral resolution, blending, etc. The fitting of a model
spectrum to the {\it entire} feature is expected to overcome these difficulties,
provided the underlying model is not too far from reality. 

\section{Comparing the results from different methods}\label{sec_results}

Using {\tt SYNOW} as described above, we determined the best-fitting parameters of all SNe spectra from Sec.~\ref{sec_data}. 
The resulting model velocities are collected in Table \ref{vel} in Appendix B. The best-fitting {\tt SYNOW} parameters,
such as $\tau_{ref}$ for each atom/ion, the power-law exponent $n$ and $v_{model}$ together with the chosen $T_{phot}$, 
can be found in Table \ref{synowtable} in Appendix C. In Table \ref{vel} we also list the
$v_{Fe}$ and $v_{H\beta}$ velocities. 
For SNe 1999em, 2005cs and 2006bp, we collected the photospheric velocities from $\tt CMFGEN$ models 
of \citet{dessart2006} and \citet{dessart2008}. These are included in Table \ref{vel} as $v_{nlte}$. 
Velocities from the cross-correlation technique (Sect. \ref{sec_cross}) were obtained using two sets of 
template spectra. The first set contained the 22 observed spectra of SN~1999em (set \#1), 
while the second set was based on the $\tt CMFGEN$ models mentioned above (set \#2). 
The velocities of the template spectra were $v_{Fe}$ for set \#1 and $v_{nlte}$ for set \#2. 
We cross-correlated all the observed spectra with the two sets separately on the wavelength range of 4500 -- 5500 \AA,
and the resulting velocities are also in Table \ref{vel} as $v_{cc}$. 

Fig.~\ref{velocities} shows $v_{model}$ against phase for all studied SNe (top left panel), 
and the ratio of $v_{model}$ to all the other velocities.
The calculated velocities all show the expected decline with phase as the photosphere moves
deeper and deeper within the ejectra, toward slower expanding layers. 

Similar plots containing the ratio $v_{model} / v_{cc}$ and $v_{abs} / v_{cc}$ as functions of phase, 
are presented in Fig.~\ref{velcc}. 

In the followings we provide some details of deriving these
velocities for each object and discuss some object-specific differences between them.

\subsection{SN 1999em}\label{sec_results99em}

\begin{figure*}
\begin{center}
\includegraphics[width=0.85\textwidth]{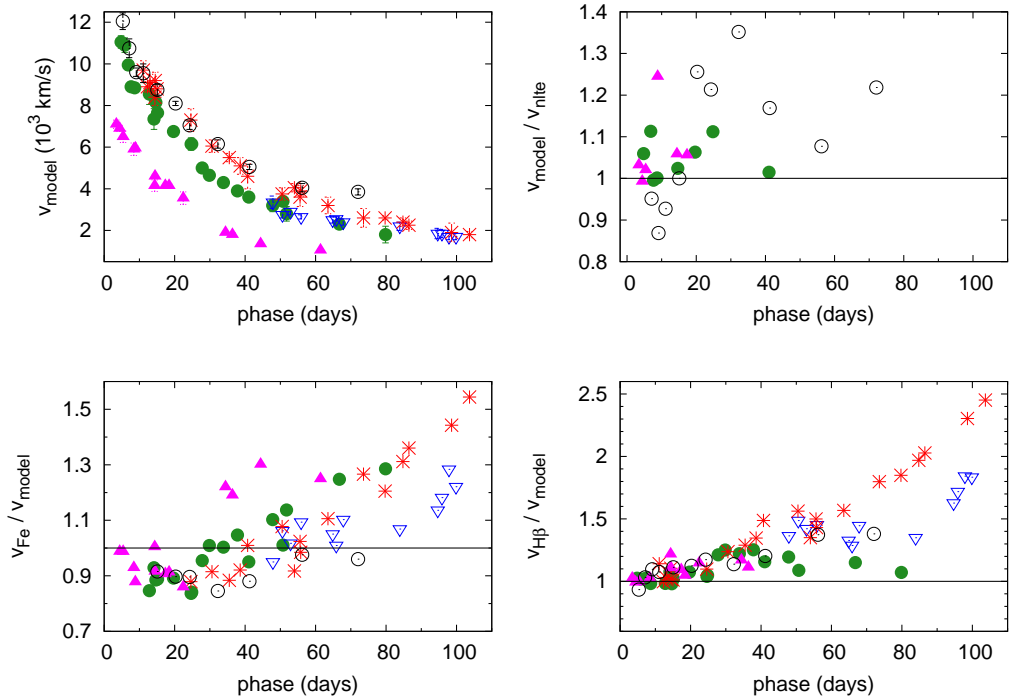}
\caption{Top left panel: model velocities ($v_{model}$) from {\tt SYNOW} as functions of phase. 
Top right panel: the ratio of {\tt SYNOW} and {\tt CMFGEN} model velocities ($v_{model} / v_{nlte}$) against phase.
Bottom panels: The phase dependence of $v_{Fe} / v_{model}$ (bottom left) and $v_{H\beta} / v_{model}$ (bottom right).
Different symbols code the followings: filled circles -- SN~1999em; open circles -- SN~2006bp; filled triangles -- SN~2005cs;
open triangles --  SN~2004dj; asterisks -- SN~2004et.}
\label{velocities}
\end{center}
\end{figure*}

When determining $v_{model}$ with {\tt SYNOW}, $H\beta$ was fitted for the first 6 spectra,
then the Fe\,{\sc ii} $\lambda5169$ feature was used for the remaining 16 spectra.
The resulting velocities are between $11000$ and $1800$ km s$^{-1}$. As seen in the bottom right panel of Fig. \ref{velocities}, 
$v_{model}$ and $v_{H\beta}$ are about the same for the early phases (before the appearance of the Fe\,{\sc ii} lines), while 
later $v_{H\beta}$ tends to be higher than $v_{model}$. Also, between day +15 and day +40, $v_{model}$ is a slightly higher than 
$v_{Fe}$ (Fig. \ref{velocities} bottom left panel). After day +40 $v_{model}$ drops below $v_{Fe}$ and their ratio increases toward later phases.

\begin{figure*}
\begin{center}
\includegraphics[width=0.85\textwidth]{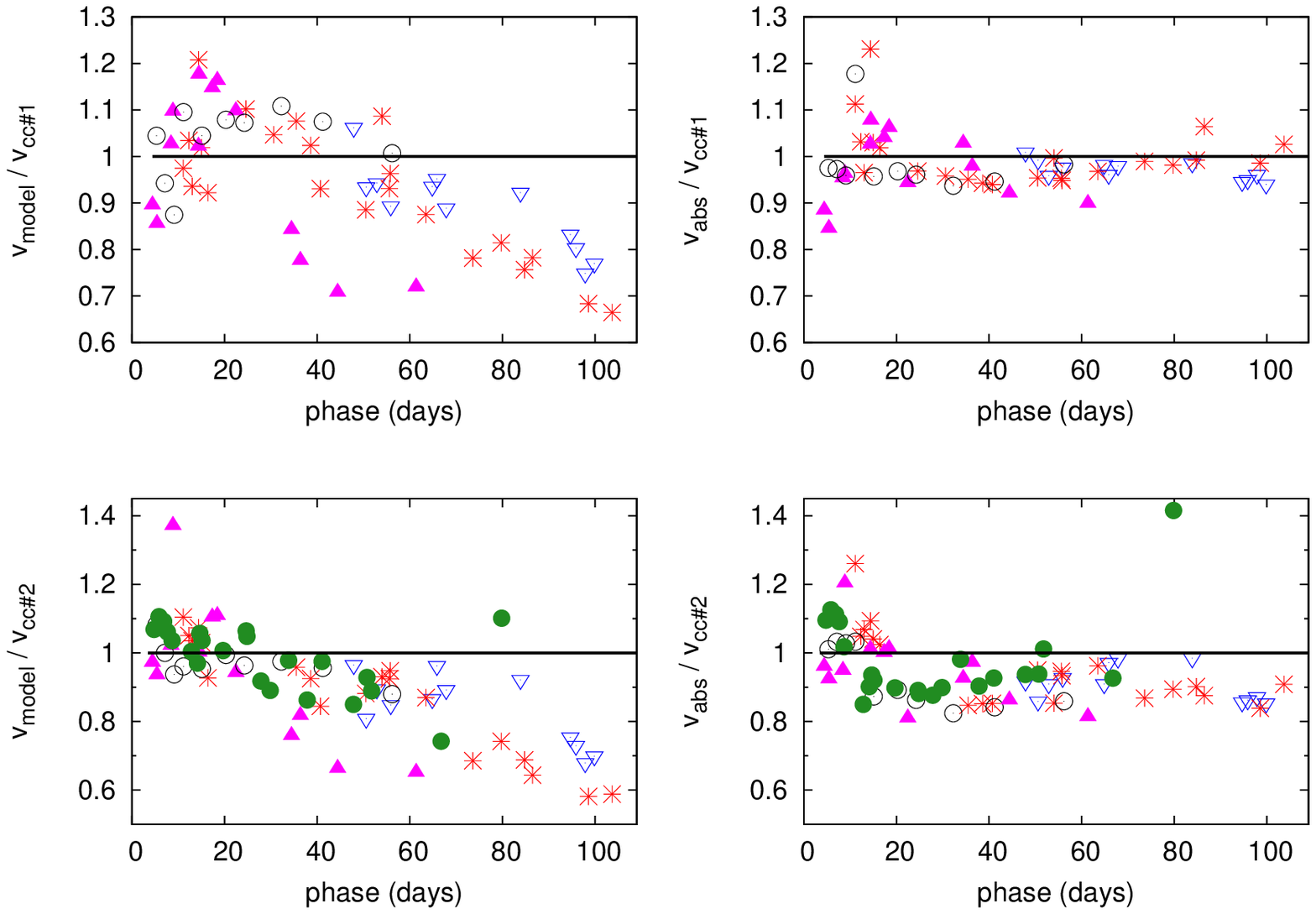}
\caption{Top panels: Ratio of $v_{model}$ and $v_{abs}$ to $v_{cc}$ from cross-correlating with the observed spectra of SN 1999em (set \#1, see text) as functions of phase.
Bottom panels: the same as above, but with respect to the template set containing CMFGEN models (set \#2). The symbols code the same SNe as in Fig.\ref{velocities}.}
\label{velcc}
\end{center}
\end{figure*}

The velocities from {\tt CMFGEN} models of \citep{dessart2006} (Fig. \ref{velocities} top right panel) agree with $v_{model}$. The cross-correlation with set \#2 (Fig.\ref{velcc} bottom panels) gave similar results for the first few points, but overestimate $v_{model}$ between days +22 and +80. They mostly fall between $v_{H\beta}$ and $v_{Fe}$, which is expected, since we cross-correlated the range of 4500 -- 5500 \AA, where these features appear.

\subsection{SN 2004dj}\label{sec_results04dj}

The {\tt SYNOW} model velocities of the 11 spectra that cover the second half of the plateau phase are between $\sim$ 3400 and 
1700 km s$^{-1}$. These are similar to those of SN 1999em at the same phase. 
Both $v_{Fe}$ and $v_{H\beta}$ are higher than $v_{model}$ at all epochs,  especially the latter with a factor 
of about $1.8$ (Fig. \ref{velocities}).  

No {\tt CMFGEN} model was available for SN~2004dj.  Cross-correlation with both template sets gave very similar results. They are only slightly higher than both $v_{model}$ and $v_{Fe}$ (Fig.\ref{velcc}).

\subsection{SN 2004et}\label{sec_results04et}

For the first 6 spectra the {\tt SYNOW} model was optimized for $H\beta$, then
for the Fe\,{\sc ii} $\lambda 5169$ feature.  
The resulting model velocities are between 9700 and 1800 km s$^{-1}$ (Fig. \ref{velocities}). 
The $v_{Fe}$ values are similar to $v_{model}$, but their ratio shows slight phase dependence, similar to the other SNe studied here.
On the contrary, the values of $v_{H\beta}$ are very different from $v_{model}$. At early phases they are close to $v_{model}$
(except for the first point), but later the $v_{H\beta}$ to $v_{model}$ ratio strongly increases, becoming as high as $2.5$.

Again, there is no {\tt CMFGEN} model available for this SN. Cross-correlation with set \#1 resulted 
in velocities similar to $v_{H\beta}$ at early phases and to $v_{Fe}$ later. With set \#2, cross-correlation gave similar results at early phases, but later it produced systematically higher velocities. This underlines the importance of selecting proper template spectra 
and template velocities when applying the cross-correlation technique.

\subsection{SN 2005cs}\label{sec_results05cs}

We used the H$\beta$ line for fitting the first 3 spectra with {\tt SYNOW}. The velocities of this SN are very low: they are in the range of 7100 -- 1100 km s$^{-1}$ and decrease quickly. The velocities from absorption minima are very close to $v_{model}$ 
for both H$\beta$ and Fe\,{\sc ii} $\lambda 5169$. The $v_{Fe}$ values follow the tendency similar to the previous objects: they are somewhat lower than $v_{model}$ at the early phases, but get higher after about day +30. 
The $v_{H\beta}$ values are much closer to $v_{model}$ than for the other SNe, and the
$v_{H\beta} / v_{model}$ ratio stays about the same for all epochs (Fig. \ref{velocities}).
The velocities of the CMFGEN models for SN~2005cs are the same as the $v_{model}$ values for all epochs, except for day +9. Cross-correlation with both template sets resulted in velocities close to $v_{Fe}$. 

\subsection{SN 2006bp}\label{sec_results06bp}

The results for SN 2006bp are controversial. Applying {\tt SYNOW}, the H$\beta$ line was fitted for the first 4 of the 11 observed spectra,
while Fe\,{\sc ii} $\lambda 5169$ was used for the rest. The model velocities are between 12000 and 3800 km s$^{-1}$. Both $v_{Fe}$ and $v_{H\beta}$ follow the tendency shown by other SNe (Fig. \ref{velocities}).

On the contrary, unlike in the previous two cases, the velocities from the {\tt CMFGEN} models differ significantly from our $v_{model}$ values. At early epochs this difference is much lower ($\sim 500 - 700$ km s$^{-1}$) being close to zero at day +15. After day +15 it gets higher reaching $\sim 1200$ km s$^{-1}$ on day +32. At later phases the difference decreases somewhat, but stays being significant. 

Cross-correlating the same spectra with the CMFGEN models using the wavelength range of $4500 - 5500$ \AA\  resulted in velocities that are very close to $v_{model}$ (except for day +9). Using the \#1 template set, the results agree well with $v_{Fe}$, or $v_{H\beta}$ at early epochs. 

\begin{figure*}
\begin{center}
\includegraphics[width=0.8\textwidth]{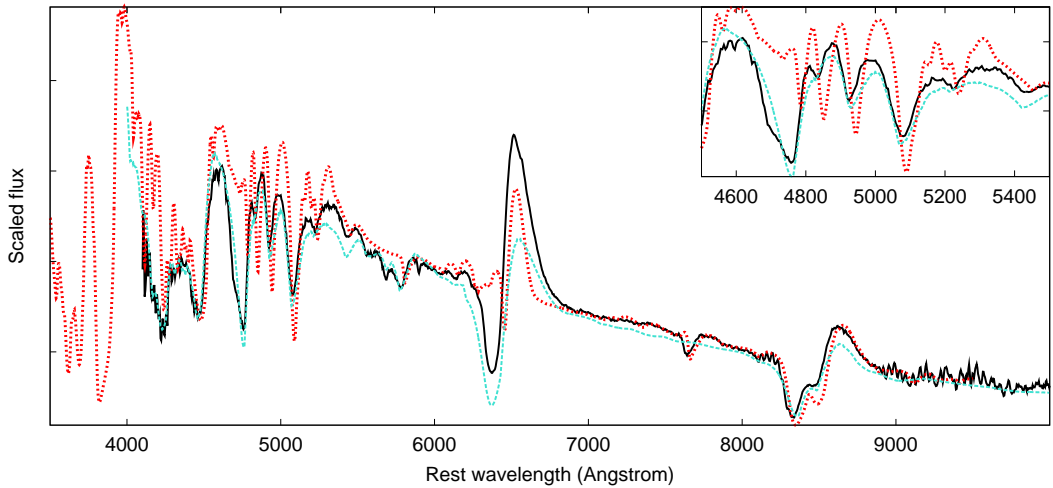}
\caption{Plot of the CMFGEN (red dotted line) and the SYNOW models (turquoise dashed line) and the observed spectra (black continuous line) of SN 2006bp on day $+$ 32 (see text).}
\label{06bp}
\end{center}
\end{figure*}

To examine the obvious controversy between the velocity of the {\tt CMFGEN} models and all the others, we plotted the observed spectra and the best-fitting {\tt CMFGEN} model on day +32 (when the differences are the highest) in Fig.~\ref{06bp}. Zooming in on the range of $4500 - 5500$ \AA~ clearly shows that the model by \citet{dessart2008} does not fit these spectral features well, leading to an underestimate of the velocity. Thus, we suspect that the velocity differences we found are probably due to the inferior fitting of the {\tt CMFGEN} models to the SN~2006bp spectra.

\section{Discussion}\label{sec_disc}

\begin{figure*}
\begin{center}
\subfigure{
\includegraphics[width=0.48\textwidth]{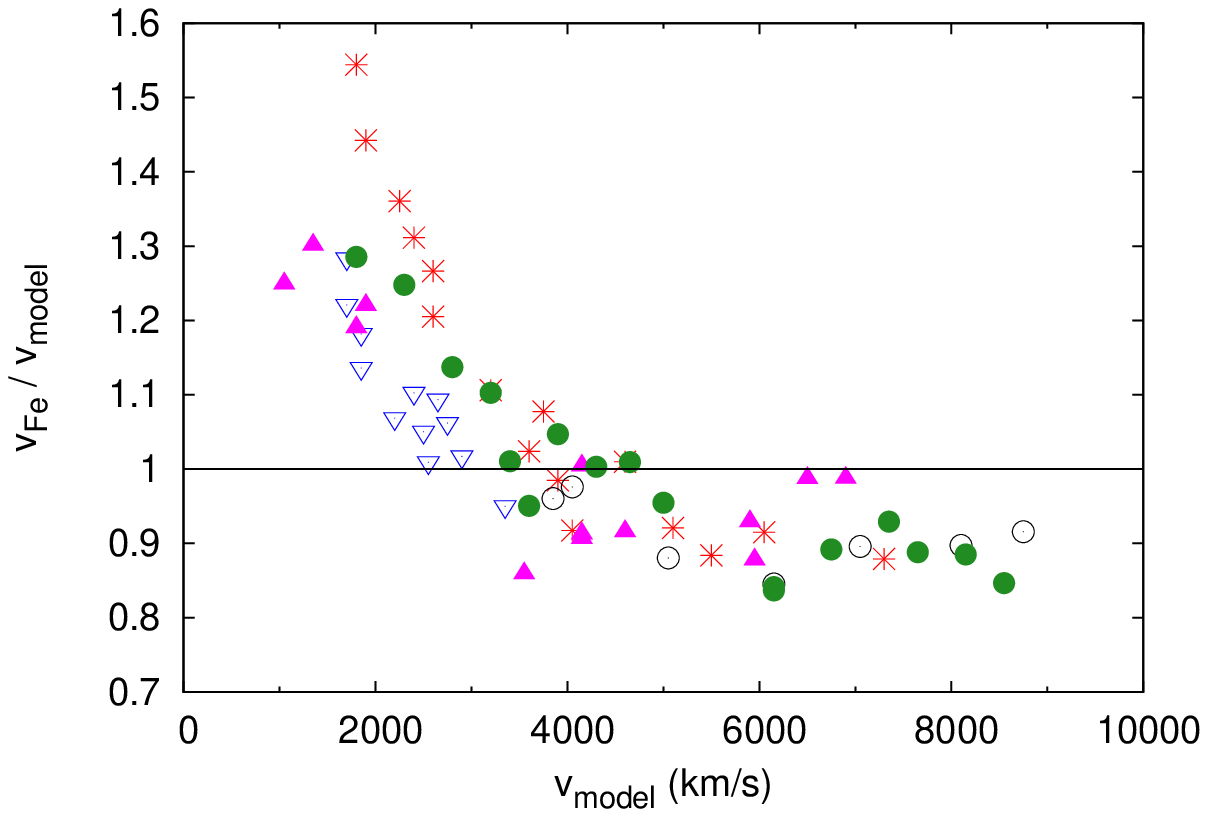} }
\subfigure{
\includegraphics[width=0.48\textwidth]{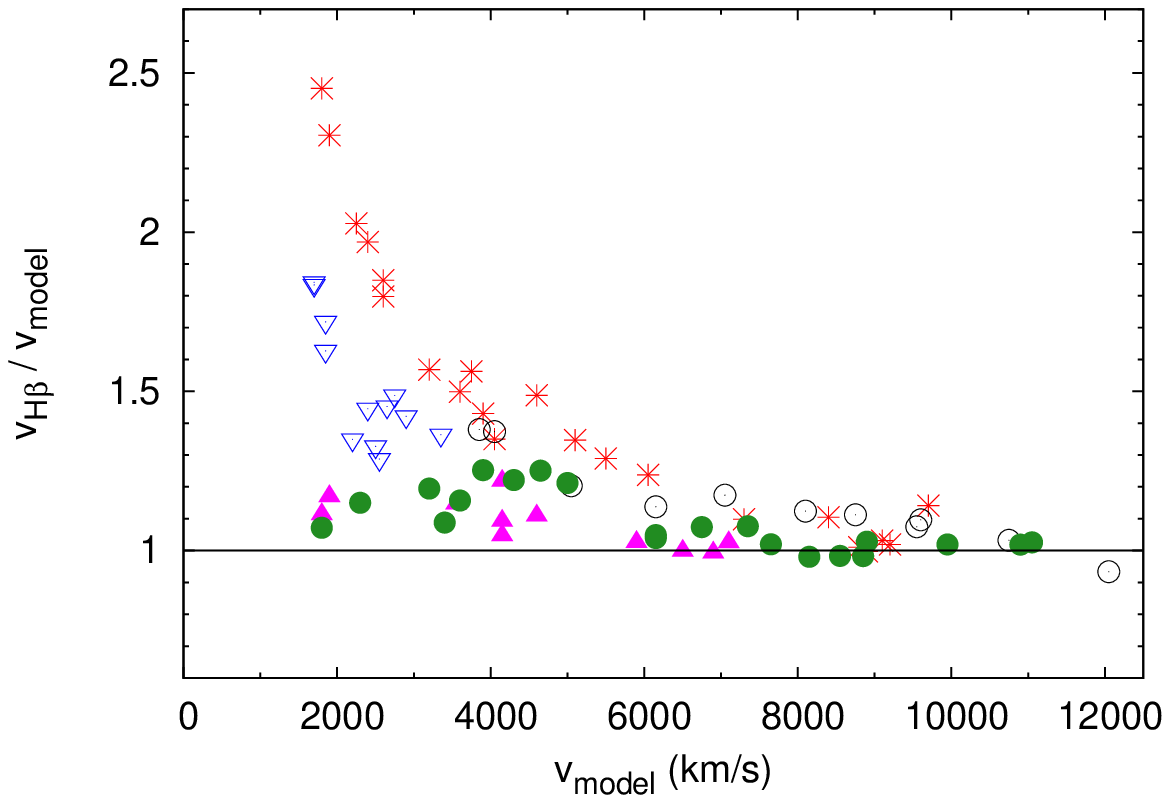}}
\caption{The $v_{Fe}/v_{model}$ (left panel) and $v_{H\beta} / v_{model}$ (right panel) ratio against the model velocities. 
The symbols have the same meaning as in Fig.\ref{velocities}. }
\label{allsn}
\end{center}
\end{figure*}

As shown in the previous sections, 
the photospheric velocities of four SNe in our sample evolved similarly. SNe 1999em, 2004et and 2006bp had high velocities at early phases and they decreased quickly, although their decline slopes were different. SN 2004dj probably showed similar evolution, but the lack of the early-phase data prevents a more detailed comparison. On the contrary, SN 2005cs was a very different, low-energy SN II-P as discussed in detail in previous studies. It had lower early velocities and the velocity curve decreased much faster than for all the other SNe. 

As expected, the different velocity measurement methods we applied provided somewhat different results. 
As seen in Fig.\ref{velcc}, the velocities obtained from cross-correlation are usually closer to $v_{abs}$ than to $v_{model}$. 
This is understandable, given that the cross-correlation method is most sensitive to the shapes and positions of the spectral 
features that may be biased toward lower, or higher velocities. The $v_{model} / v_{cc}$ ratio (Fig.\ref{velcc} top left 
and bottom left panels) shows the same trend (but plotted upside down) as the $v_{Fe} / v_{model}$ ratio in 
Fig.\ref{velocities} (bottom left panel), i.e. $v_{model}$ is higher between day 10 and 50, but becoming smaller 
than $v_{Fe}$ or $v_{cc}$. On the other hand, no such systematic trend can be identified between $v_{model}$ and
$v_{nlte}$ (Fig.\ref{velocities}, top right panel). These benchmarks suggest that the model velocities, either
from {\tt SYNOW} or {\tt CMFGEN} are consistent, and they show phase-dependent offsets from the absorption minima,
or cross-correlation velocities. The increasing systematic offset is particulary strong for $v_{H\beta}$ (Fig.\ref{velocities}
bottom right panel). Thus, the traditional, simple measurement methods seem to underestimate the true photospheric
velocities before day 50, but increasingly overestimate them toward later epochs. This should be kept in mind
when the true photospheric velocities are needed, e.g. in the application of EPM.

In order to do further testing, we plotted the ratio of $v_{Fe} / v_{model}$ and $v_{H\beta}/v_{model}$ as a function 
of $v_{model}$ for all SNe, following \citet{dessart2005b} (Fig. \ref{allsn}).  
The $v_{Fe}/v_{model}$ ratio shows the same trend for all objects: at high velocities (i.e. early phases) the ratio is somewhat lower than 1, then it reaches unity around $\sim 4000$ km s$^{-1}$, and below that it keeps rising, reaching $\sim$ 1.6 by the end of the plateau. The $v_{H\beta} / v_{model}$ ratio is more complicated. At high $v_{model}$ values it is around 1, but becomes higher than unity around $v_{model} \approx 7000$ km s$^{-1}$. Below that the slope of the rising changes from object to object. In the case of SNe 2005cs and 1999em this ratio stays under $\sim 1.4$, while for the other three SNe it becomes much higher. For SN 2006bp there are no spectra below $v_{model} = 3850$ km s$^{-1}$, but above that its evolution seems to be similar to that of SN 2004et.

A similar plot was published by \citet{dessart2005b} based on their set of {\tt CMFGEN} model spectra (see their Fig. 14). 
The only slight difference is
that they plotted the ratio of the velocity measured from the absorption minima of the model spectra to the input velocity of the code, as a function of the input velocities. Although they did not have data below $\sim$ 4000 km s$^{-1}$, and we do not have data 
above $\sim$ 12050 km s$^{-1}$, between these limits their plotted values are mostly similar to ours. In their Fig.14 the 
Fe\,{\sc ii} $\lambda$5169 velocities are lower than that of the model for high velocities, and their ratio reaches 1 between 5000 and 4000 km s$^{-1}$, just like our data. The situation is somewhat different for H$\beta$. At high velocities the two results are consistent: above $\sim 11000$ km s$^{-1}$ the data by \citet{dessart2005b}, as well as ours, are around 1. However, their velocity ratio exceeds 1 at $\sim 8000$ km s$^{-1}$ and has a highest value of 1.15 for H$\beta$. It is much lower than our results in Fig.\ref{allsn}. In the case of SN 2004et, our velocity ratio goes as high as 2.5. It must be noted, however, that the model spectra used by 
\citet{dessart2005b} were tailored to represent SNe 1987A and 1999em (\citeauthor{dessart2005a}). The latter object is also in our sample, and our $v_{H\beta}$ to $v_{model}$ ratio for that particular SN is similar to the results of \citet{dessart2005b}. Thus, it is probable that the lower $v_{H\beta} / v_{model}$ ratio of \citet{dessart2005b}
is due to the limited parameter range of their {\tt CMFGEN} models used to create their plot.

Recently \citet{roy} published a study of velocity measurement for the Type II-P SN 2008gz. They applied a similar technique of using {\tt SYNOW} to fit Fe\,{\sc ii} features around 5000 \AA. They also estimated the velocity from the absorption minima of these lines. They got $4200 \pm 400$ km s$^{-1}$ and $4000 \pm 300$ km s$^{-1}$ for $v_{model}$ and $v_{Fe}$, respectively, from a $+$87d spectrum. This result is consistent with our findings plotted in Fig. \ref{allsn}: $v_{Fe}$ is practically equal to $v_{model}$ around $4000$ km s$^{-1}$.

\begin{figure*}
\subfigure[]{
\includegraphics[width=84mm]{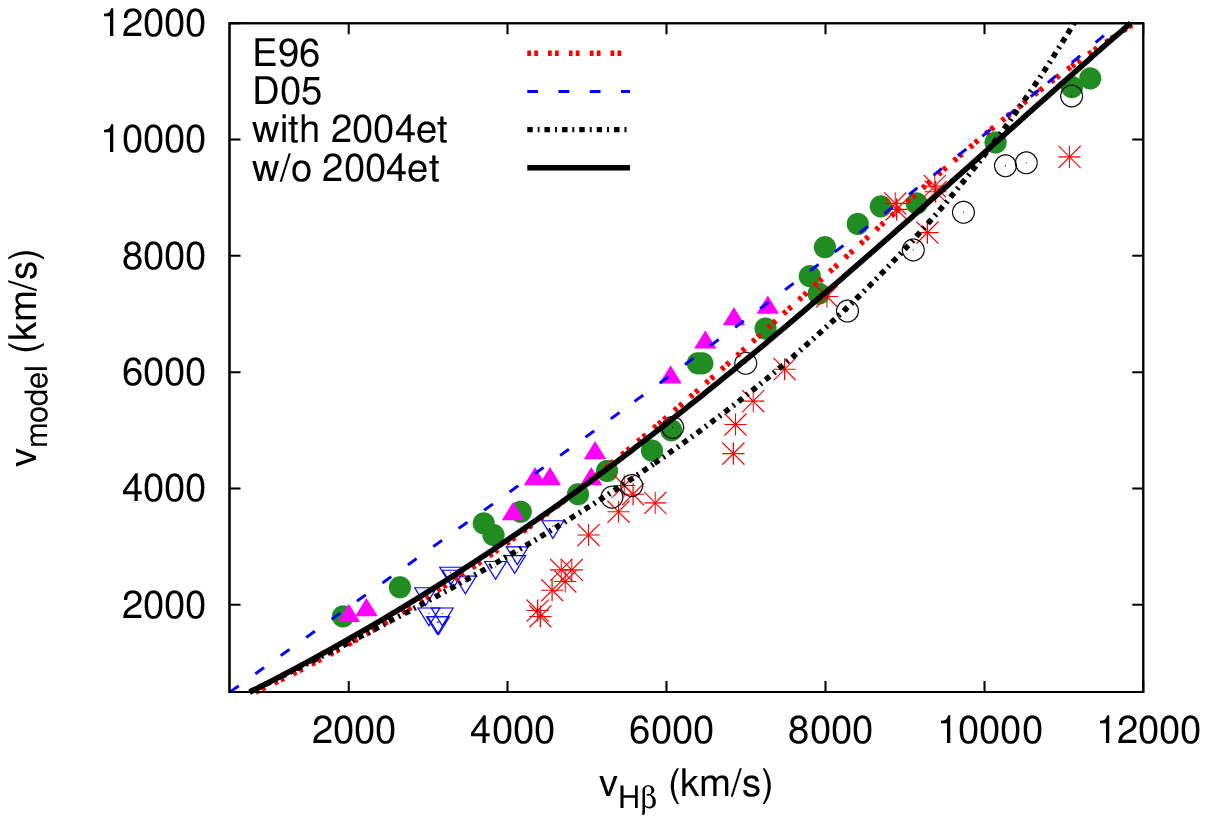}
\label{j-h-a}}
\subfigure[]{
\includegraphics[width=84mm]{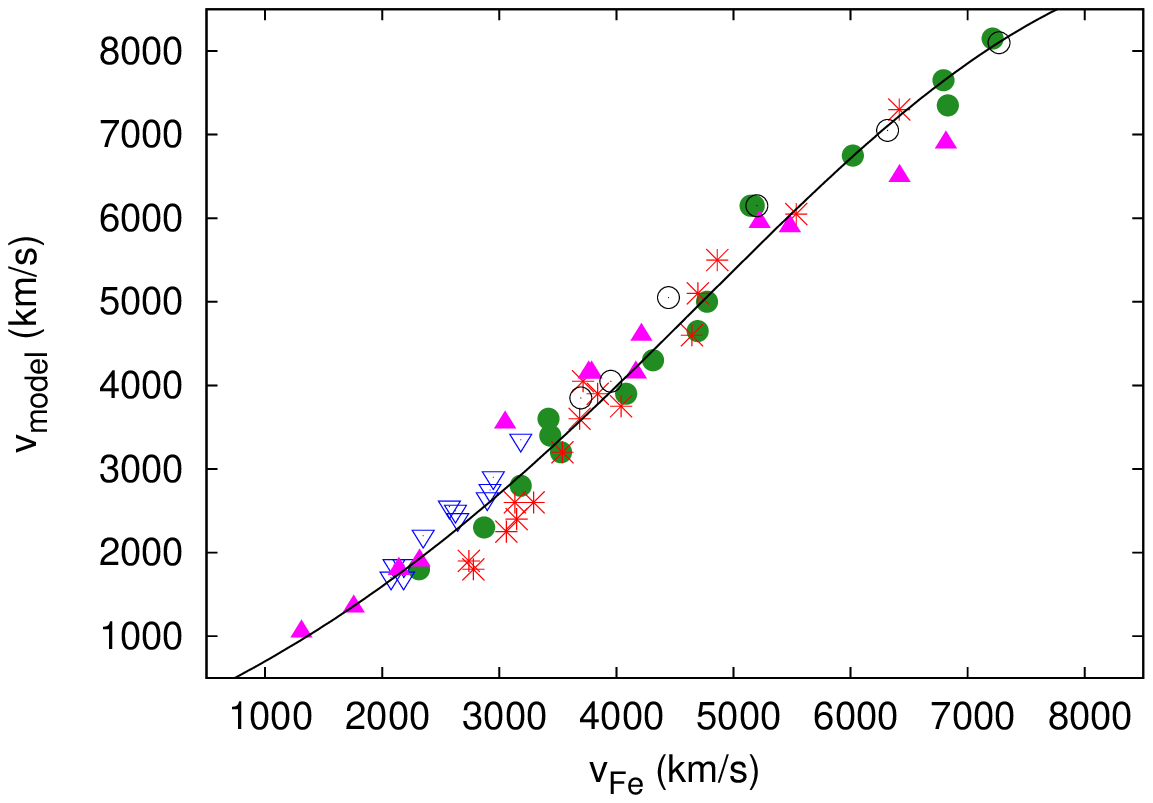}
\label{j-h-b}}
\caption{(a) The plot of $v_{model}$ against $v_{H\beta}$. The lines represent the polynomials calculated by \citet{joneshamuy} (see text) based on the models of \citeauthor{E96} (red dotted line), \citeauthor{dessart2005a} (blue dashed line). The result from our fitting on all SNe is plotted as a black dotted line line while the fitting omitting SN 2004et is shown by a black solid line. (b) The same fitting as (a) but for $v_{Fe}$ using all SNe. The simbols have the same meaning as in Fig.\ref{velocities}. The polynomial coefficients are in Table \ref{jhtable}.}
\label{j-h}
\end{figure*}

\subsection{Velocity-velocity relations}\label{vel-vel}

Using the synthetic spectra of \citeauthor{E96} and \citeauthor{dessart2005a}, \citet{joneshamuy} also examined the relation between $v_{H\beta}$ and $v_{model}$. They found that their ratio can be described as
\begin{equation}
{ v_{H\beta} \over v_{model}} = \sum_{j=0}^2 a_j v_{H\beta}^j
\label{eq1}
\end{equation}
where the values of $a_j$ are given in Table \ref{jhtable}. In Fig. \ref{j-h-a} we plotted our data together with these polynomials. The polynomials based on the \citeauthor{dessart2005a} models overestimate our $v_{model}$ values 
(rms $\sigma=0.412$), while those from the E96 models provide much better fit for all SNe except SN~2004et
(rms $\sigma=0.301$, but $\sigma=0.178$ without the data of SN 2004et).  

We fitted Eq.(\ref{eq1}) to our data (Fig. \ref{j-h-a}, black curve).  The resulting $a_j$ coefficients are in Table \ref{jhtable}. Our fit resulted in a much lower rms scatter, $\sigma=0.276$.
Repeating the fitting while omitting the data of SN 2004et, the result became very similar to that from the E96 models.

Since $v_{Fe}$ is thought to be a better representative of the velocity at the photosphere than $v_{H\beta}$, it is expected that $v_{model}$ can be predicted with better accuracy by measuring $v_{Fe}$.
Indeed, Fig. \ref{allsn} suggests that the $v_{Fe}/v_{model}$ ratio is almost the same from SN to SN, unlike the $v_{H\beta}/v_{model}$ ratio that can be quite different for different SNe. Thus, we repeated the fitting of Eq.(\ref{eq1}) using $v_{Fe}$ instead of $v_{H\beta}$ (Fig. \ref{j-h-b}).  We found the rms scatter of $\sigma=0.111$, which is much lower than in the previous cases. The $a_j$ coefficients of this fitting are also included in Table \ref{jhtable}. 

The tight relation between $v_{Fe}$ and $v_{model}$ in Fig. \ref{j-h-b} suggests a possibility to 
{\it estimate} $v_{model}$ from the measured $v_{Fe}$ values. However, it is emphasized that
SN-specific differences in the expansion velocities may exist, thus, model building 
for a particular SN, whenever possible, should always be preferred.

\begin{figure*}
\subfigure[]{
\includegraphics[width=84mm]{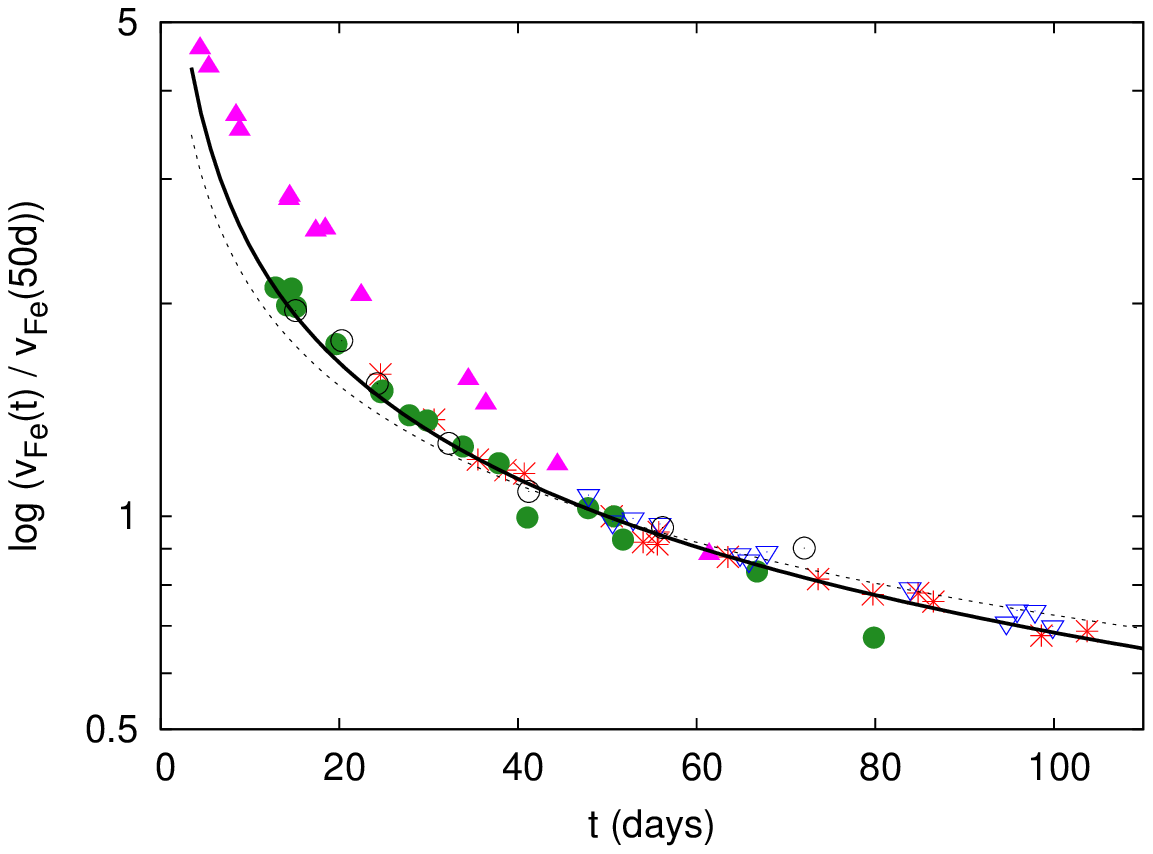}
\label{nug-a}}
\subfigure[]{
\includegraphics[width=84mm]{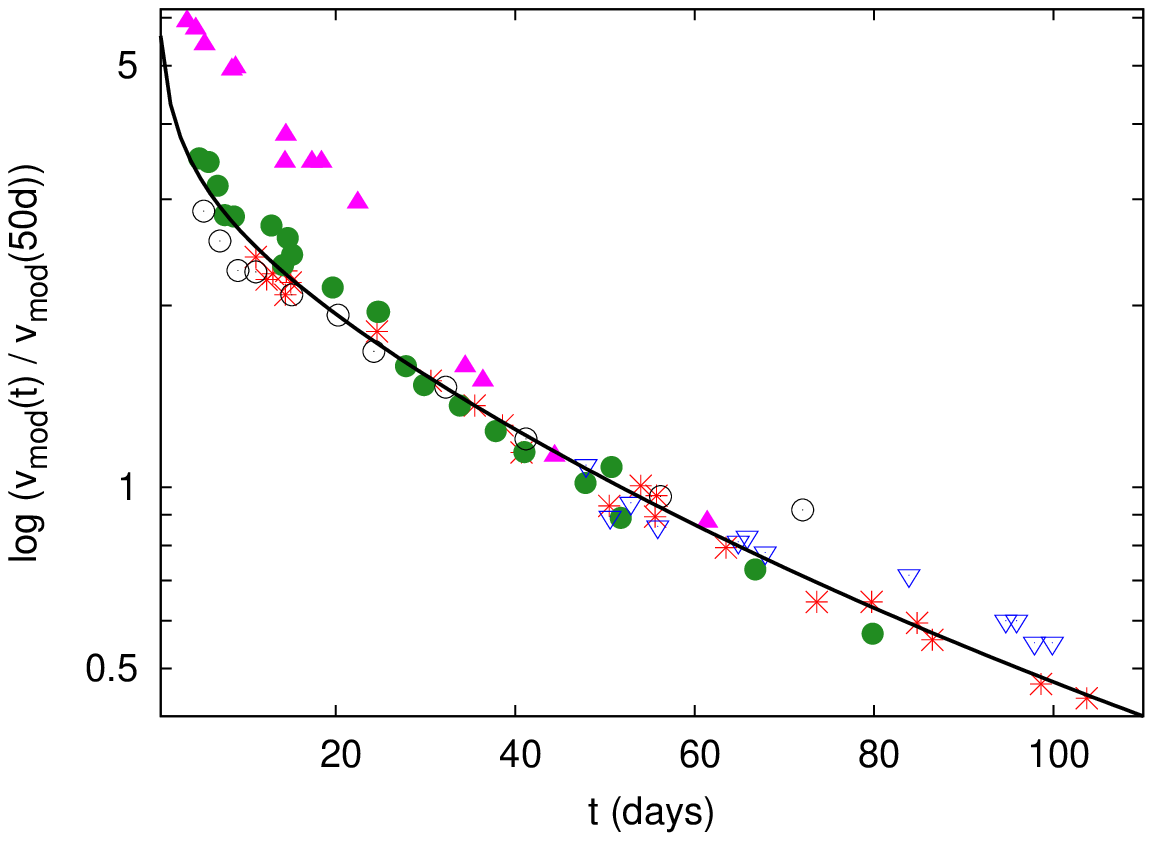}
\label{nug-b}}
\caption{(a) The evolution of $v_{Fe}(t)/v_{Fe}(50\rmn{d})$ and (b) the same for $v_{model}(t)/v_{model}(50\rmn{d})$. The dashed line is the result by \citet{nugent06}, while the solid lines are from our fittings (see text). The data of SN 2005cs (filled triangles) were excluded from the fitting. The color codes are the same as in Fig. \ref{allsn}.}
\label{nug}
\end{figure*}

\citet{nugent06} found that $v_{Fe}$ evolves as

\begin{equation}
 v_{Fe}(t)/v_{Fe}(50\rmn{d})=(t/50)^{c}
\label{eq2}
\end{equation}
 where $c=-0.464 \pm 0.017$. After repeating the fitting of Eq.(\ref{eq2}) to our data, we found the exponent to be $c=-0.663 \pm 0.01$. Then, since the data of SN~2005cs are very different from the rest of the sample, we omitted the velocities of SN~2005cs and repeated the fitting.  This resulted in $c= -0.546 \pm 0.01$ (Fig. \ref{nug-a}). 
These two exponents marginally differ (at $\sim 1 \sigma$) from the value given by \citet{nugent06}. A possible source of this difference (beside the different velocity measurement techniques applied) may be that our sample covers the phases between +13 and +104 days, while the data by \citet{nugent06} are between +9 and +75 days.

We also examined how the {\tt SYNOW} model velocities evolve in time. Combining Eq.(\ref{eq1}) and Eq.(\ref{eq2}), the following relation has be derived (again, excluding SN 2005cs from the sample):
\begin{equation}
v_{model}(t) / v_{model}(50\rmn{d})= {(t/50)^{-0.210 \pm 0.11} \over \sum_{j=0}^2 b_j(t/50)^j }
\label{eq3}
\end{equation}
where $b_0=0.467 \pm 0.15$, $b_1=0.327 \pm 0.23$ and $b_2=0.174 \pm 0.11$. The rms scatter is $\sigma=0.148$ (Fig. \ref{nug-b}).

\begin{table}
 \centering
 \begin{minipage}{84mm}
  \caption{Polynomial coefficients for the $v_{H\beta}$ and $v_{Fe}$ to $v_{model}$ ratio (Eq.\ref{eq1}).}
  \label{jhtable}
  \begin{tabular}{@{}lccccl@{}}
  \hline
  j & 0 & 1 & 2 & $\sigma$ & ref. \\
 \hline
 $a_j$ ($H_{\beta}$) (E96) & 1.775 & -1.435e-4  & 6.523e-9 & 0.30 & (1) \\
 $a_j$ ($H_{\beta}$) (D05) & 1.014 & 4.764e-6 & -7.015e-10 & 0.41 & (1) \\
 $a_j$ ($H_{\beta}$) & 1.528 & -1.551e-5 & -3.462e-9 & 0.27 & (2) \\
 $a_j$ ($H_{\beta}$) w/o 04et & 1.578 & -8.573e-5 & 3.017e-9 & 0.17 & (2) \\
 $a_j$ (Fe\,{\sc ii} $\lambda5169$) & 1.641 & -2.297e-4 & 1.751e-8 & 0.11 & (2) \\
\hline
\end{tabular}
 \begin{tablenotes}
       \item[a]{(1) \citet{joneshamuy}}
       \item[b]{(2) this paper}
     \end{tablenotes}
\end{minipage}
\end{table}

\begin{figure}
\includegraphics[width=84mm]{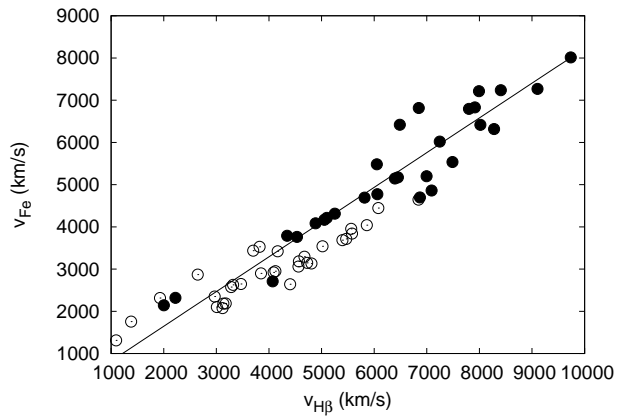}
\caption{The relation between the measured $v_{Fe}$ and $v_{H\beta}$ values. The empty circles show the data taken after day 40, while the filled circles refer to data taken before day 40. The line shows the fit to the filled circles only.}
\label{pozn}
\end{figure}

As was mentioned in \S \ref{sec_pcyg}, using SDSS data \citet{poznanski} examined the correlation between velocities measured from the absorption minima of H$\beta$ and Fe\,{\sc ii} $\lambda 5169$ lines (see Fig. \ref{pozn}). They found that there is a linear relation given by $v_{Fe}(50\rmn{d})=a \cdot v_{H\beta}(50\rmn{d})$, where $a=0.84 \pm 0.05$. Using our sample we repeated their fitting. First, we used all epochs where both $v_{H\beta}$ and $v_{Fe}$ were measured. The slope of the fitted line was $a=0.791 \pm 0.012$ ($\sigma=0.146$). Then, we kept only the velocities obtained before day 40 \citep[similar to][]{poznanski}. This resulted in $a=0.823 \pm 0.015$ ($\sigma=0.102$), which is basically the same as that of \citet{poznanski}. Thus, our study fully confirms the results by \citet{poznanski}, but
extends the validity of the relation toward later phases.  
 
\section{Implications for distance measurements}\label{sec_dist}

Improving the accuracy of the velocity measurements has an important aspect in measuring extragalactic distances with SCM or EPM (cf. \S 1). Both SCM and EPM need velocities, thus, the results of this paper can be significant for both techniques.

SCM was calibrated using $v_{Fe}$ on day +50. However, in many cases getting a spectrum at or around day +50 is not possible. In these cases Eq.(\ref{eq2}) can be used to estimate $v_{Fe}(50\rmn{d})$. We have improved the exponent in Eq.(\ref{eq2}) as $-0.546 \pm 0.01$, based on more data  obtained on wider range of phase than previously. The difference between our result and the previous curve \citep{nugent06} is the highest around day +20 (Fig.\ref{nug}). The new curve may result in better constrained $v_{Fe}(50\rmn{d})$ when only early-phase spectra obtained around day +20 are available. 

However, there are several drawbacks of SCM. For example, the uncertainty in the moment of explosion, i.e. in determining the phase of a particular spectrum can lead to significant error in the distance determination. Moreover, as the example of SN 2005cs shows in Fig.\ref{nug}, some Type II-P SNe can deviate significantly from the average, especially during early phases. Thus, one should be careful when such kind of interpolation or extrapolation is to be applied. The example of SN 2005cs suggests that multi-epoch spectroscopic observations should always be preferred against single-epoch spectra when distance determination is the aim.

The case of EPM is different. Since this method does not require calibration, but needs multi-epoch data, deviations in the measured 
velocities have higher impact. To show this, we calculated the EPM--distances of all 5 SNe via the method described in \citet{vinko11dh}. We used two sets of velocities for each SNe: $i)$ from the absorption minimum of the Fe $\lambda$5169 line, $ii)$ $v_{model}$ determined in 
Sec. \ref{sec_synow}.

The resulted distances are in Table \ref{epm}. The correction factors of \citeauthor{dessart2005a} were applied for all SNe. Usually the photometric data were interpolated to the epochs of the velocities. 
However, for SN 2004dj Eq. \ref{eq2} and Eq. \ref{eq3} were used to extrapolate the velocity data to the photometric epochs, because 
of the low number of spectra taken before day +50, i.e. during the expansion of the photosphere. 

\subsection {SN 1999em}\label{sec_dist99em}

Data on SN 1999em in NGC~1637 were used for distance determination with EPM several times. \citet{hamuy2001} used cross-correlation  velocities (with the model spectra of E96 as the template set) and the correction factors of E96 to obtain the distance of $7.8 \pm 0.5$ Mpc. Using absorption minima velocities and the same correction factors, \citet{leonard99em} determined the distance as $8.2 \pm 0.6$ Mpc, while \citet{elmhamdi99em} obtained $7.8 \pm 0.3$ Mpc. On the other hand, \citet{leonard_ceph} determined the distance of NGC 1637 using Cepheids as $11.7 \pm 1$ Mpc, 
which is significantly higher. Using the SEAM method, \citet{baron99em} got $12.5 \pm 1.8$ Mpc. With the velocities of their CMFGEN models and and the correction factors of D05, \citet{dessart2006} derived $11.5 \pm 1$ Mpc, and with a similar approach to that of \citet{baron99em} they obtained $12.2 \pm 2$ Mpc. Recently \citet{joneshamuy} estimated the photospheric velocity from $v_{H\beta}$ (\S \ref{vel-vel} Eq. \ref{eq1}), and derived $9.3 \pm 0.5$ Mpc by using the correction factors of E96, and $13.9 \pm 1.4$ Mpc from the correction factors of D05.

We have repeated the EPM analysis using D05 correction factors and our $v_{model}$ velocities.  This resulted in $12.5 \pm 1.4$ Mpc, which is in good agreement with the cepheid- and SEAM-distances and that of \citet{dessart2006} using the velocities determined from their CMFGEN models. Instead, applying the $v_{Fe}$ velocities (extrapolating for the first few points using Eq.\ref{eq2}), the distance became lower, $9.7 \pm 0.8$ Mpc. The disagreement between these two distances is roughly the same as that due to the application of different
correction factors (see above). The distance obtained from adopting the $v_{model}$ velocities is in much better agreement
with the independent Cepheid-based distance to the host galaxy. It suggests that the application of the proper velocity data
is important to obtain more realistic and bias-free distances from Type II-P SNe.

\subsection{SN 2004dj}\label{sec_dist04dj}

There are many published distances for the host galaxy of SN 2004dj (NGC 2403), but they show large scatter, being 
between $2.88$ and $6.43$ Mpc, according to the NED\footnote{http://ned.ipac.caltech.edu/} database.  

In the case of this SN the velocity curve was extrapolated using Eq. \ref{eq2} and Eq. \ref{eq3} to the epochs of the photometry of \citet{vinko04dj} and \citet{tsvetkov04dj}. The distances from the two velocity curves agree very well ($D=3.6 \pm 0.6$ and $3.7 \pm 0.8$ Mpc, respectively, Table \ref{epm}), and they are also in very good agreement with the result of \citet{vinko04dj}.

\subsection{SN 2004et}\label{sec_dist04et}

Similarly, the distances of NGC 6946, the host galaxy of SN 2004et, show large scatter being in the range of $4.7$ and $7.2$ Mpc (NED). 
$D \sim 4.7$ Mpc was derived recently by \citet{poznanski09} using SCM.

Our result supports this shorter value. The EPM-analysis with both $v_{model}$ and $v_{Fe}$ (after extrapolation to the early phases by Eq. \ref{eq2}) gave $4.8$ Mpc (Table \ref{epm}).

\subsection{SN 2005cs}\label{sec_dist05cs}

In the case of SN 2005cs the application of $v_{model}$ resulted in a considerably longer distance ($8.6 \pm 0.2$ Mpc) than the one 
using $v_{Fe}$ ($7.5 \pm 0.2$ Mpc). This longer distance is in good agreement with a recent study by Vink\' o et al. (2011), in preparation, who determined the distance of M51 via EPM by combining the data of SNe 2011dh and 2005cs, and obtained $8.4 \pm 0.7$ Mpc. The reason for the slight difference between their result and ours with $v_{model}$ (although both papers used the same photometry, velocities and method) is that, unlike  \citet{vinko11dh}, we did not fix the moment of explosion in EPM. Instead, we also optimized that parameter to keep consistency with the analysis of all the other objects in this paper.

\subsection{SN 2006bp}\label{sec_dist06bp}

For SN 2006bp and its host galaxy, NGC 3953, the distances  are between $15.7$ and $21.0$ Mpc (NED). Both of our results fit into this wide range, but with the usage of $v_{model}$ we obtained slightly longer values ($20.7 \pm 1.8$ Mpc) than with $v_{Fe}$ ($18.6 \pm 1.5$ Mpc), the latter being closer to the distance of \citet{dessart2008}, i.e. $17.1$ and $17.5$ Mpc from SEAM and SCM, respectively.

\begin{table}
 \centering
 
  \caption{The EPM distances of the 5 SNe using different velocities.}
  \label{epm}
  \begin{tabular}{@{}lccc@{}}
  \hline
  SN & \multicolumn{2}{c}{D (Mpc)} &  Photometry\footnotemark[1] \\
& with $v_{model}$ & with $v_{Fe}$ & \\
 \hline
1999em &  12.5 (1.4)  & 9.7 (0.8)  &  1,2,3 \\
2004dj & 3.6 (0.6) & 3.7 (0.8)  &  4,5 \\
2004et & 4.8 (0.4) &  4.8 (0.6) &  6,7 \\
2005cs & 8.6 (0.2) &  7.5 (0.2)  &  8 \\
2006bp & 20.7 (1.8)  &  18.6 (1.5) & 9\\
 \hline
\end{tabular}
 \begin{tablenotes}
       \item[1]{$^1$ Source of photometry: (1) \citet{hamuy99em}, (2) \citet{leonard99em}, (3) \citet{elmhamdi99em}, (4) \citet{vinko04dj}, (5) \citet{tsvetkov04dj}, (6) \citet{maguire04et}, (6) \citet{sahu04et},   (8) \citet{pastorello05csII}, (9) \citet{dessart2008}}
       
     \end{tablenotes}

\end{table}

\section{Conclusions}\label{sec_concl}

In this paper we investigated three methods for estimating photospheric velocities of Type II-P SNe. We focused on building model spectra with {\tt SYNOW}, and compared the resulting $v_{model}$ velocities with those obtained by cross-correlation or simply measuring absorption minima of P Cygni features. Based on a sample of 81 spectra from 5 SNe, we showed that {\tt SYNOW } provides very similar photospheric velocities to those derived by more sophisticated modeling codes, but in a faster, less computation-intensive way. This approach may be more extensively applicable, yet it preserves the advantages of using physically consistent model spectra to estimate parameters of SNe non-interactively, and without relying mostly on eye-ball estimates and human decisions.  

We illustrated that the cross-correlation- and absorption minimum velocities, i.e. those determined by more
conventional methods, suffer from phase-dependent systematic deviations from the model velocities. 
This has already been known from previous studies \citep[e.g.][]{dessart2005b}, but we have extended
the phase coverage of the modeled spectra, and revealed that such deviations become stronger
below $v_{phot} \sim 3000$ km ~s$^{-1}$, i.e. after day +60. At these late phases $v_{Fe}$ may
overestimate $v_{model}$ by 30 - 50 \% depending on the atmospheric properties of the particular SN.

Based on these results, we verified and updated the relations between the photospheric velocities and the ones estimated from the Doppler-shifts of the absorption minima of individual spectral lines. It was found that while the $v_{Fe}/v_{model}$ ratio appears to be nearly the same for all SNe studied here, it is not true for the velocities from the H$\beta$ line. We have derived
a power-law relation to estimate $v_{model}$ from $v_{Fe}$ and/or $v_{H\beta}$, but due to the possibility of 
SN-dependent systematic deviations, we recommend the computation of parametrized models, whenever possible.  

Using the model velocities, we re-determined the distances of the 5 SNe via EPM, and compared them with the ones calculated by using 
$v_{Fe}$. The distances obtained from $v_{model}$ are similar or slightly higher than those with $v_{Fe}$.  For SN~1999em, which is the most thoroughly studied object in our sample, we were able to show that by using the model velocities the derived distance is more consistent with the Cepheid-based distance to the host galaxy. Although such a comparison was
not possible for the other SNe due to the lack of reliable Cepheid distances, this result underlines the importance of the
velocity measurement method in SN distance studies. 

Despite its numerous advantages, EPM also suffers from caveats. One of them is the need for many photometric data and contemporaneous velocities (i.e. spectra) covering most of the plateau phase. This is hardly achievable for most SNe. A possible solution may be a careful interpolation between the measured data points. Previously, the weakest link was the poorly resolved velocity curve, thus, mainly the light curves were interpolated to the
moments of velocity measurements \citep[e.g.][]{hamuythesis}. Based on our results in Sec. \ref{sec_results}, the interpolation of the velocity curve to the epochs of photometric data via Eq.(\ref{eq3}) may also be a possibility, 
resulting in a better sampled dataset for EPM. We intend to demonstrate the application of this approach for new SNe in a future paper (Tak\'ats et al., in preparation).

\section*{Acknowledgement}

This project is supported by the European Union and co-funded by
the European Social Fund through the T\' AMOP 4.2.2/B-10/1-2010-0012 grant.
This work has also been partly supported by the Hungarian OTKA Grant K76816, the Hungarian National Office of Research and Technology, 
NSF Grant AST-0707769, and Texas Advanced Research Project grant ARP-0094 for J.C. Wheeler at
University of Texas at Austin. 
 We thank Dr. A. Pastorello and Dr. K. Maguire for providing spectra of SNe 2005cs and 2004et in digital
form, and Dr. L. Dessart for sending their {\tt CMFGEN} models used in this paper. 
We are grateful to Prof. S. Rucinski, S. Mochnacki, T. Bolton and R.
Garrison (University of Toronto) for their generous offer of their telescope
time used for observing SN 2004et at DDO in 2004. We also express our thanks to the referee, Dr. D. Poznanski for his thorough report that helped us to improve the manuscript.
The NASA ADS and NED databases and the Supernova Spectrum Archive (SUSPECT) were used to access data and references. The availability of these services are gratefully acknowledged.

\appendix\label{appendix}

\section{Early spectra of SN 2004\lowercase{et}}\label{appendix_a}

\begin{figure}
\includegraphics[width=84mm]{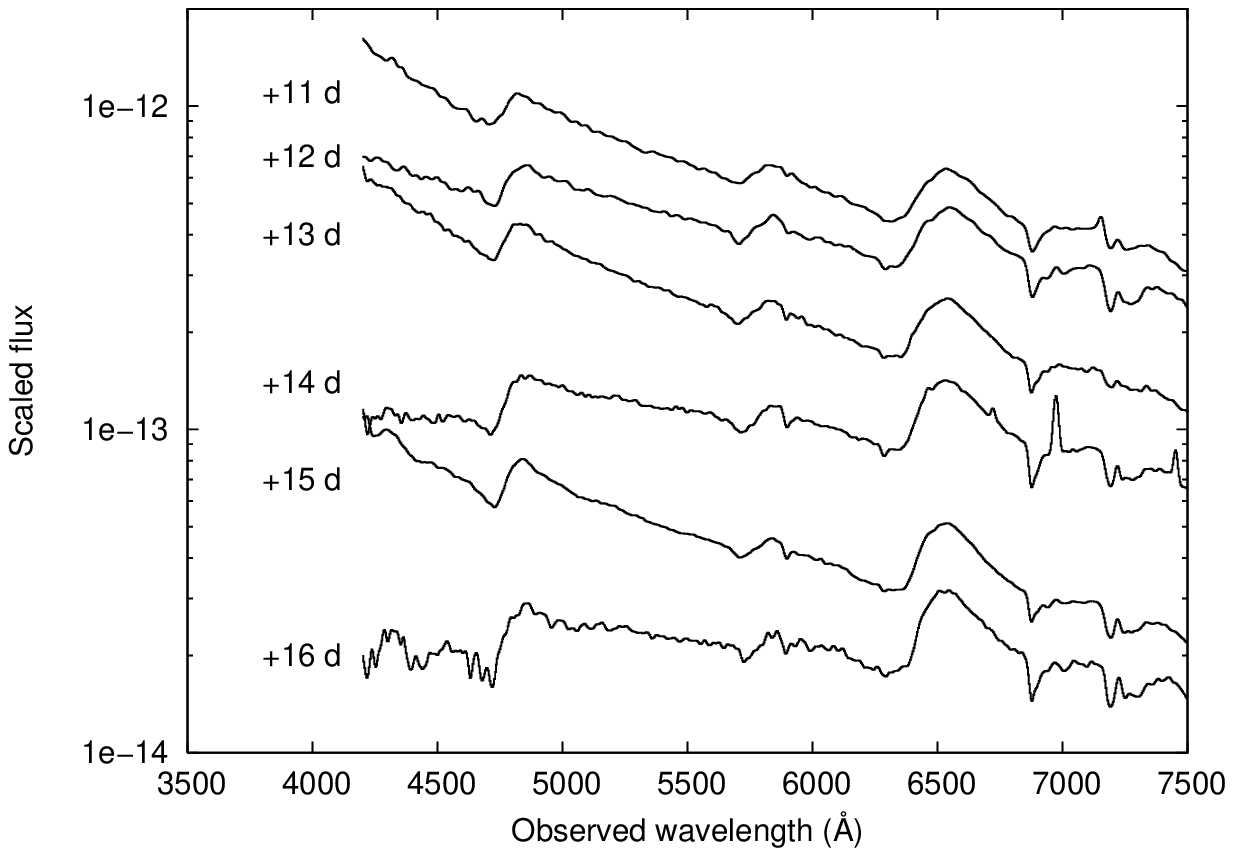}
\caption{Early spectra of SN 2004et}
\label{04etsp}
\end{figure}

Soon after its discovery \citep{discov04et}, six low-resolution 
optical spectra of SN~2004et were taken at the David Dunlap Observatory,
Canada, with the Cassegrain spectrograph mounted on the 74" telescope.
The spectra covered the 4000 - 8000 \AA\ regime with a resolution
of $\sim 800$ at 6000 \AA \citep[see the details on the data reduction in][]{vinko04dj}. 
Due to the fixed North-South slit direction, the spectra could not be taken at
the parallactic angle, thus, the slope of the continuum in the blue is affected
by differential refraction.
Table~\ref{04ettable} contains the journal of these
observations, and the spectra are plotted in Fig. \ref{04etsp}.

\begin{table}
\centering
\begin{minipage}{84mm}
\caption{Journal of spectroscopic observations of SN~2004et}
\begin{tabular}{lcccl}
\hline
Date & JD-2,450,000 & Phase(d)  & Airmass & Observer\footnotemark[1] \\
\hline
2004-10-03 & 3281.6 & +11 & 1.2 & JT, TK \\
2004-10-04 & 3282.8 & +12 & 2.0 & SM, JT \\
2004-10-05 & 3283.5 & +13 & 1.1 & HD, TK \\
2004-10-06 & 3284.9 & +14 & 2.3 & HD, JT \\
2004-10-07 & 3285.5 & +15 & 1.1 & HD, JT \\
2004-10-08 & 3286.9 & +16 & 1.7 & JT, JG \\
\hline
\end{tabular}
\begin{tablenotes}
	\item{$^1$ Observers: JT: J. Thomson, TK: T. Koktay, SM: S. Mochnacki, HD: H. DeBond, JG: J. Grunhut}
       
     \end{tablenotes}
\label{04ettable}
\end{minipage}
\end{table}

\section{SNe velocities measured with different techniques}\label{appendix_b}

In Table \ref{vel} we present the velocities obtained with {\tt SYNOW} ($v_{model}$, see \S \ref{sec_synow}), along with those measured from the absorption minima of H$\beta$ and Fe\,{\sc ii} $\lambda5169$ ($v_{H\beta}$ and $v_{Fe}$), as well as those obtained with cross-correlation method using the observed spectra of SN 1999em as templates ($v_{cc\#1}$) and the CMFGEN models ($v_{cc\#2}$).

\renewcommand{\thefootnote}{\alph{footnote}}
\begin{table*}
 \centering
 \begin{minipage}{180mm}
 \caption{The velocities (in km s$^{-1}$) obtained with different techniques: the photospheric velocities from the best-fitting {\tt SYNOW} models ($v_{model}$), the Doppler-shift of the absorption minima of H$\beta$ and Fe\,{\sc ii} $\lambda5169$ features ($v_{H\beta}$ and $v_{Fe}$, respectively) and the cross-correlation velocities from templates consisting of the observed spectra of SN 1999em ($v_{cc\#1}$) 
and the {\tt CMFGEN} models ($v_{cc\#2}$). The phase (in days) were calculated relative to the date in Table \ref{physdata}.}
  \label{vel}
  \smallskip 
  \addtocounter{footnote}{1}
\begin{tabular}{@{}ccccccccccccc@{}}
\hline
\multicolumn{6}{c}{SN 1999em} & & \multicolumn{6}{c}{SN 2004et} \\
  \hline
  phase & $v_{model}$ & $v_{H\beta}$ & $v_{Fe}$ & $v_{cc\#1}$  & $v_{cc\#2}$ && phase & $v_{model}$ & $v_{H\beta}$ & $v_{Fe}$  & $v_{cc\#1}$ & $v_{cc\#2}$ \\
\hline  
4.79	 &  11050 (300) & 11332 & -- &   &  10341 (454) && 11.10 & 9700 (450) & 11072 & --  & 9951 (153) &  8781 (1316) \\
5.84	 &  10900 (350) & 11101 & -- &   &  9858  (888) && 12.30 & 8900 (250) & 8878 & --   & 8605 (86)  &  8470 (681) \\ 
6.84	 &  9950 (250) & 10141 & --  &   &  9112  (878) && 13.00 & 9100 (400) & 9386 & --   & 9724 (117) &  8786 (1099) \\ 
7.64	 &  8900 (200) & 9148 & --   &   &  8383  (832) && 14.40 & 9200 (400) & 9375 & --   & 7617 (71)  &  8570 (482) \\ 
8.67	 &  8850 (150) & 8697 & --   &   &  8544  (491) && 15.00 & 8800 (100) & 8894 & --   & 8635 (64)  &  8545 (524) \\ 
12.84	 &  8550 (500) & 8406 & 7236 &   &  8513  (497) && 16.40 & 8400 (400) & 9282 & --   & 9110 (73)  &  9060 (545) \\ 
14.14	 &  7350 (500) & 7913 & 6829 &   &  7575  (289) && 24.60 & 7300 (550) & 8018 & 6416 & 6624 (49)  &  --  \\
14.67	 &  8150 (350) & 7992 & 7213 &   &  7708  (190) && 30.60 & 6050 (300) & 7487 & 5535 & 5779 (56)  &  --  \\
15.14	 &  7650 (150) & 7802 & 6793 &   &  7380  (187) && 35.50 & 5500 (200) & 7091 & 4861 & 5112 (43)  &  5739 (350) \\
19.67	 &  6750 (200) & 7246 & 6019 &   &  6703  (242) && 38.60 & 5100 (400) & 6869 & 4695 & 4981 (45)  &  5513 (337) \\
24.66	 &  6150 (300) & 6393 & 5146 &   &  5779  (126) && 40.70 & 4600 (600) & 6842 & 4644 & 4943 (41)  &  5448 (256) \\
24.84	 &  6150 (225) & 6450 & 5172 &   &  5866  (194) && 50.50 & 3750 (300) & 5859 & 4040 & 4236 (49)  &  4252 (367) \\
27.84	 &  5000 (200) & 6060 & 4773 &   &  5446  (155) && 54.00 & 4050 (200) & 5468 & 3715 & 3727 (23)  &  4352 (244) \\
29.84	 &  4650 (250) & 5816 & 4693 &   &  5221  (201) && 55.60 & 3600 (450) & 5396 & 3686 & 3867 (37)  &  3895 (408) \\
33.84	 &  4300 (100) & 5251 & 4312 &   &  4394  (215) && 55.76 & 3900 (250) & 5578 & 3840 & 4048 (37)  &  4113 (228) \\
37.84	 &  3900 (100) & 4885 & 4083 &   &  4521  (159) && 63.50 & 3200 (400) & 5018 & 3539 & 3656 (43)  &  3678 (349)  \\
41.04	 &  3600 (200) & 4164 & 3421 &   &  3691  (115) && 73.60 & 2600 (450) & 4674 & 3292 & 3327 (45)  &  3793 (217) \\
47.84	 &  3200 (400) & 3822 & 3528 &   &  3765  (163) && 79.73 & 2600 (100) & 4807 & 3133 & 3192 (31)  &  3504 (246) \\
50.74	 &  3400 (150) & 3699 & 3435 &   &  3661  (131) && 84.79 & 2400 (200) & 4725 & 3147 & 3172 (51)  &  3490 (257) \\
51.76	 &  2800 (350) & --   & 3183 &   &  3147  (152) && 86.50 & 2250 (200) & 4561 & 3060 & 2876 (64)  &  3496 (148) \\
66.76	 &  2300 (100) & 2645 & 2869 &   &  3097  (146) && 98.60 & 1900 (450) & 4377 & 2740 & 2779 (22)  &  3268 (131) \\
79.84	 &  1800 (400) & 1928 & 2313 &   &  1634  (100) && 103.7 & 1800 (150) & 4412 & 2779 & 2708 (33)  &  3060 (201) \\

\hline	   		
\multicolumn{6}{c}{SN 2004dj} && \multicolumn{6}{c}{SN 2005cs} \\
  \hline
  phase & $v_{model}$ & $v_{H\beta}$ & $v_{Fe}$ & $v_{cc\#1}$  & $v_{cc\#2}$ && phase & $v_{model}$ & $v_{H\beta}$ & $v_{Fe}$ & $v_{cc\#1}$ & $v_{cc\#2}$ \\
\hline 
 47.89 & 3350 (300) & 4569 & 3183  & 3154 (40) & 3471 (83)   && 3.44 &  7100 (200) & 7275 & --    & 17105 (1876) & 30045 (365) \\
 50.59 & 2750 (150) & 4089 & 2920  & 2943 (63) & 3401 (172)  && 4.41 &  6900 (150) & 6848 & 6814  & 7703  (109)  & 7092 (705) \\
 52.89 & 2900 (150) & 4122 & 2949  & 3075 (75) & 3243 (115)  && 5.39 &  6500 (250) & 6487 & 6418  & 7594  (151)  & 6945 (586) \\
 55.89 & 2650 (100) & 3849 & 2897  & 2967 (72) & 3124 (122)  && 8.43 &  5900 (300) & 6050 & 5481  & 5745  (143)  & 5775 (71) \\
 64.86 & 2500 (150) & 3317 & 2625  & 2672 (45) & 2887 (114)  && 8.84 &  5950 (100) & --   & 5222  & 5422  (209)  & 4337 (193) \\
 65.85 & 2550 (100) & 3282 & 2573  & 2678 (47) & 2649 (106)  && 14.36 & 4150 (300) & 5052 & 4167  & 4060  (52)	& 4119 (80) \\
 67.86 & 2400 (150) & 3468 & 2645  & 2700 (46) & 2689 (118)  && 14.44 & 4600 (100) & 5099 & 4213  & 3908  (43)	& 4591 (170) \\
 83.89 & 2200 (200) & 2967 & 2350  & 2384 (50) & 2387 (114)  && 17.35 & 4150 (50)  & 4532 & 3763  & 3616  (49)	& 3757 (129) \\
 94.67 & 1850 (250) & 3010 & 2101  & 2222 (67) & 2454 (145)  && 18.42 & 4150 (100) & 4345 & 3788  & 3566  (58)	& 3742 (110) \\
 95.87 & 1850 (200) & 3178 & 2186  & 2302 (71) & 2535 (161)  && 22.45 & 3550 (300) & 4069 & 3049  & 3232  (51)	& 3765 (315) \\
 97.87 & 1700 (150) & 3132 & 2182  & 2270 (78) & 2504 (156)  && 34.44 & 1900 (50)  & 2222 & 2318  & 2254  (19)	& 2505 (126) \\
 99.86 & 1700 (150) & 3118 & 2075  & 2208 (86) & 2433 (153)  && 36.40 & 1800 (100) & 2003 & 2142  & 2189  (38)	& 2202 (74) \\
       &            &      &       &           &             && 44.40 & 1350 (50)  & --   & 1756  & 1907  (23)	& 2036 (182)  \\
       &            &      &       &           &             && 61.40 & 1050 (50)  & --   & 1311  & 1459  (24)	& 1611 (274) \\
\hline
\multicolumn{13}{c}{SN 2006bp} \\
\hline
  phase & $v_{model}$ & $v_{H\beta}$ & $v_{Fe}$ & $v_{cc\#1}$  & $v_{cc\#2}$ && phase & $v_{model}$ & $v_{H\beta}$ & $v_{Fe}$ & $v_{cc\#1}$ & $v_{cc\#2}$ \\
\hline
5.30  & 12050 (400) & 11251 & --  &  11537 (138) & 11131 (571) && 24.26 & 7050  (200) & 8278 & 6315  & 6573 (45)   & 7317  (329)\\
7.10  & 10750 (450) & 11097 & --  &  11407 (103) & 10748 (875) && 32.26 & 6150  (200) & 6996 & 5199  & 5548 (28)   & 6308  (330)\\
9.10  & 9600  (200) & 10526 & --   & 10975(70)	 & 10242 (657) && 41.21 & 5050  (125) & 6078 & 4445  & 4698 (46)   & 5281  (181)\\
11.11 & 9550  (450) & 10265 & --   & 8717 (69)	 & 9941  (521) && 56.19 & 4050  (200) & 5562 & 3952  & 4022 (40)   & 4602  (180)\\
15.10 & 8750  (250) & 9735 & 8013  & 8373 (67)	 & 9176  (325) && 72.04 & 3850 (150)    & 5313   & 3697   & --		  &  -- \\
20.28 & 8100  (100) & 9104 & 7267  & 7506 (79)	 & 8147  (184) &&       &            &       &        &            & \\
\hline
\end{tabular}						                				 
\end{minipage}  					                				 
\end{table*}

\section{Parameters of the best-fitting SYNOW models}\label{appendix_c}

In Table \ref{synowtable} we present the parameters of the best-fitting {\tt SYNOW} models, including: $\tau_{ref}$ of the main atoms/ions, the power-law exponent of the optical depth function ($n$), the photospheric temperature ($T_{bb}$) and photospheric 
velocity ($v_{phot}$) (see \S \ref{sec_synow} for details).

\renewcommand{\thefootnote}{\alph{footnote}}
\begin{table*}
 \centering
 \begin{minipage}{168mm}
 \caption{The {\tt SYNOW} parameters of the best-fitting models of SN 1999em. The phases were calculated from the assumed
moments of explosions listed in Table \ref{physdata}.}
  \label{synowtable}
  \smallskip 
  \addtocounter{footnote}{1}
\begin{tabular}{@{}ccccccccccccccc@{}}
  \hline
\multicolumn{15}{c}{SN 1999em}  \\
\hline 
phase & JD-2,451,000 &  \multicolumn{10}{c}{$\tau_{ref}$} & n & $T_{bb}$ & $v_{phot}$   \\
  (days) & (days)  & \mbox{H\,{\sc i}} & \mbox{He\,{\sc i}}  & \mbox{Na\,{\sc i}}  & \mbox{Si\,{\sc i}} & \mbox{Si\,{\sc ii}} & \mbox{Ca\,{\sc ii}} & \mbox{Sc\,{\sc ii}} & \mbox{Ti\,{\sc ii}} & \mbox{Fe\,{\sc ii}} & \mbox{Ba\,{\sc ii}} & & (kK) & (km s$^{-1}$) \\
 \hline
 4.79\footnotemark[1]  & 481.79  & 2.80 & 0.25 && &&&& &&& 3.0 & 14.2 & 11050   \\
 5.84\footnotemark[2]  & 482.84  & 3.50 & 0.20 & &&&&& &&& 3.0 & 12.0 & 10900    \\
 6.84\footnotemark[2]  & 483.84  & 4.90 & 0.40 & &&&&& &&& 3.0 & 11.0 & 9950   \\
 7.64\footnotemark[1]  & 484.64  & 6.30 & 0.35 & &&&&& &&& 3.0 & 9.5 & 8900   \\
 8.67\footnotemark[1]  & 485.67  & 7.30 & 0.20  &&&& &&& && 3.5 & 13.6 & 8850   \\
 12.84\footnotemark[2] & 489.84  & 15.80 & 0.10  &&&&& && && 5.5 & 10.0 & 8550  \\
 14.14\footnotemark[2] & 491.14  & 21.10 & &  0.20 & && &&& 0.30&& 5.0 & 11.0 & 7350  \\
 14.67\footnotemark[1] & 491.67  & 26.15 & 0.05 &  0.10 &&& 2.20 & & & 0.80 & 0.35 & 6.5 & 11.5 & 8150   \\
 15.14\footnotemark[1] & 492.14  & 20.20 & &  0.10 & & 0.25 & & & & 0.70 & & 4.5 & 10.4 & 7650   \\
 19.67\footnotemark[1] & 496.67  & 42.05 & &  0.10 & & & 16.0 & 0.05 & 0.30 & 0.95 & 0.40 & 5.0 & 9.0 & 6750   \\
 24.66\footnotemark[1] & 501.66  & 113.50 & & 0.55 &&& 131.9 & 0.25 & 2.10 & 1.85 & & 8.0 & 8.3 & 6150  \\
 24.84\footnotemark[2] & 501.84  & 81.00 & & 0.35 & &&17.6 & 0.15 & 1.10 & 1.60 & 0.40 & 7.0 & 8.3 & 6150   \\
 27.84\footnotemark[2] & 504.84  & 73.95 & & 0.40 & &&& 0.30 & 1.80 & 2.25 & 0.20 & 6.0 & 8.2 & 5000   \\
 29.84\footnotemark[2] & 506.84  & 65.00 & &  0.45 &  && & 0.15 & 1.65 & 4.10& & 6.0 & 7.0 & 4650   \\
 33.84\footnotemark[2] & 510.84  & 42.20 & & 1.10 & 0.01 && 95.0 & 0.05 & 1.20 & 2.20 & & 4.5 & 5.2 & 4300   \\
 37.84\footnotemark[2] & 514.84  & 39.60 && 1.00& 0.01 & 0.10 & & 0.15 & 1.55 & 2.25 & & 4.5 & 5.5 & 3900   \\
 41.04\footnotemark[2] & 518.04  & 57.20 &  0.20 & 1.70&0.01 &0.35 & 458.0 & 0.20 & 4.05 & 4.25 & 0.20 & 7.0 & 6.8 & 3600  \\
 47.84\footnotemark[2] & 524.84  & 23.00 &  & 1.85 & 0.01 & 0.35 & 200.0 & 0.45 & 2.30 & 2.20 & 1.00 & 3.0 & 4.5 & 3200  \\
 50.74\footnotemark[1] & 527.74  & 30.80 & &  2.05 & 0.01 & 0.35 & 90.0 & 0.65 & 3.75 & 3.35 & & 4.5 & 5.3 & 3400  \\
 51.76\footnotemark[1] & 528.76  & 36.60 &  & 1.95  &0.02 & & 212.0 & 0.20 & 0.50 & 2.65 & 0.25 & 3.0 & 6.5 & 2800   \\
 66.76\footnotemark[1] & 543.76  & 16.80 &  & 5.65 & 0.03& 0.40 & & 1.25 & 7.45 & 6.45 & & 3.5 & 5.5 & 2300   \\
 79.84\footnotemark[2] & 556.84  & 71.50 &  & 13.95 & 0.02 & 0.65 & 969.0 & 2.35 & 17.95 & 13.90 & 4.70 & 3.5 & 6.0 & 1800  \\
\hline
 \multicolumn{15}{c}{SN 2004dj} \\
\hline 
phase & JD-2,450,000 &  \multicolumn{10}{c}{$\tau_{ref}$} & n & $T_{bb}$ & $v_{phot}$   \\
  (days) & (days)  & \mbox{H\,{\sc i}} & \mbox{He\,{\sc i}}  & \mbox{Na\,{\sc i}}  & \mbox{Si\,{\sc i}} & \mbox{Si\,{\sc ii}} & \mbox{Ca\,{\sc ii}} & \mbox{Sc\,{\sc ii}} & \mbox{Ti\,{\sc ii}} & \mbox{Fe\,{\sc ii}} & \mbox{Ba\,{\sc ii}} & & (kK) & (km s$^{-1}$)  \\
\hline
 47.89 & 3234.89  & 81.0 & &  0.70  & & 0.15 & & 0.30 & 2.85 & 2.55 & 0.85 & 6.0 & 7.65 & 3350   \\
 50.59 & 3237.59  & 88.0 &  & 1.10  & 0.01 & && 0.70 & 6.15 & 6.00 & 1.00 & 5.5 & 6.0 & 2750   \\
 52.89 & 3239.89  & 68.0 && 1.05  & 0.01 & 0.40 & & 0.40 & 3.75 & 3.05 & 1.00 & 5.5 & 8.0 & 2900   \\
 55.89 & 3242.89  & 51.0 && 1.15  & 0.02 &  0.20 & & 0.30 & 2.75 & 3.70 & 0.80 & 4.5 & 6.2 & 2650   \\
 64.86 & 3251.86  & 72.0 &&  1.95  & 0.01 & 0.25 && 0.75 & 4.25 & 4.95 & 0.50 & 5.5 & 8.0 & 2500  \\
 65.85 & 3252.85  & 63.0 && 2.05  & 0.01 & 0.20 && 0.50 & 4.75 & 5.15 & 2.00 & 5.5 & 6.0 & 2550   \\
 67.86 & 3254.86  & 83.0 && 2.25  & 0.02 & 0.30 && 0.70 & 5.45 & 8.55 & 0.05 & 5.5 & 6.8 & 2400  \\
 83.89 & 3270.89  & 85.0 && 7.10  & 0.01 & 0.30 && 1.00 & 6.30 & 7.55 & 1.60 & 5.5 & 8.9 & 2200   \\
 94.67 & 3281.67  & 145.0 && 18.45  & 0.01 & 0.10 && 1.00 & 7.15 & 8.80 & 3.55 & 5.5 & 9.7 & 1850   \\
 95.87 & 3282.87  & 109.0 & &  24.75  & 0.02 & 0.30 & & 1.05 & 9.55 & 10.20 & 2.45 & 6.0 & 9.4 & 1850   \\
 97.87 & 3284.87  & 185.0 && 34.30  & 0.06 & 0.40 && 1.40 & 11.70 & 14.20 & 3.00 & 5.0 & 9.5 & 1700   \\
 99.86 & 3286.86  & 120.0 && 25.0  & & & & 0.50 & 9.00 & 10.0 & 0.40 & 4.5 & 8.5 & 1700  \\
\hline
\end{tabular}
 \begin{tablenotes}
	\item{Source of spectra:}
       \item[a]{$^a$ \citet{leonard99em}}
       \item[b]{$^b$ \citet{hamuy99em}}
     \end{tablenotes}
\end{minipage} 
\end{table*}

\begin{table*}
\centering
 \begin{minipage}{168mm}
\contcaption{ for SN 2004et.}
  \smallskip 
  \addtocounter{footnote}{1}
  \begin{tabular}{@{}ccccccccccccccc@{}}
  \hline
 \multicolumn{15}{c}{SN 2004et}  \\
\hline 
phase & JD-2,451,000 &  \multicolumn{10}{c}{$\tau_{ref}$} & n & $T_{bb}$ & $v_{phot}$  \\
  (days) & (days)  & \mbox{H\,{\sc i}} & \mbox{He\,{\sc i}} &    \mbox{Na\,{\sc i}}  & \mbox{Si\,{\sc i}} & \mbox{Si\,{\sc ii}} & \mbox{Ca\,{\sc ii}} & \mbox{Sc\,{\sc ii}} & \mbox{Ti\,{\sc ii}}  & \mbox{Fe\,{\sc ii}} & \mbox{Ba\,{\sc ii}} & & (kK) & (km s$^{-1}$)  \\
\hline
  11.10\footnotemark[1] & 281.60  & 3.75 & 0.20 &&&&&&&&& 4.5 & 80.0 & 9700  \\
  12.30\footnotemark[1] & 282.80  & 4.05 & 0.20 &&&&&&&&& 3.0 & 19.0 & 8900  \\
  13.00\footnotemark[1] & 283.50  & 6.40 & 0.20 &&&&&&&&& 5.0  & 52.0 & 9100  \\
  14.40\footnotemark[1] & 284.90  & 8.00 &  & 0.30  &&&&&&&& 4.0& 9.5  & 9200 \\
  15.00\footnotemark[1] & 285.50  & 7.25 & 0.15 &&&&&&&&& 5.0 & 43.0 & 8800  \\
  16.40\footnotemark[1] & 286.90  & 14.45 & 0.25 &&&&&&&&& 4.0 & 9.0 & 8400  \\
  24.60\footnotemark[2] & 295.10  & 10.0 & &  0.10 &  & & 4.00 & & & 0.85 & 0.20 & 3.0 & 11.0 & 7300  \\
  30.60\footnotemark[2] & 301.10  & 21.0 & &  0.10 & &  0.40 & 74.4 & & 0.15 & 1.20 & & 4.5 & 7.9 & 6050  \\
  35.50\footnotemark[2] & 306.00  & 41.0 & 0.10 & 0.05& & &  97.0 && 0.85 & 2.15 &  0.05 & 5.5 & 7.4 &  5500  \\
  38.60\footnotemark[2] & 309.10  & 47.0 & &  0.25 &   0.01 & & 98.0 & 0.15 & 1.45 &  2.80 & 0.05 & 4.0 & 7.2 & 5100 \\
  40.70\footnotemark[2] & 311.20  & 36.0 & & 0.15 & &  0.05 &  95.8 & 0.25 & 0.75 &  2.05 & 0.30 & 3.5 & 6.8 &  4600  \\
  50.50\footnotemark[2] & 321.00  & 40.0 & & 0.65 &  & &   324.0 & 0.25 & 1.25 &  2.35 & 0.70 & 3.0 & 5.4 & 3750  \\
  54.00\footnotemark[3] & 324.50  & 81.7 & & 0.90 &  0.01 & 0.75 & 32.7 & 0.55 & 4.60 &  3.55 & & 6.0 & 7.6 & 4050 \\
  55.60\footnotemark[2] & 326.10  & 45.0 &  & 0.95 &  0.01 & & 45.0 & 0.45 & 3.05 &  5.05 & 0.45 & 3.5 & 6.8 & 3600 \\
 55.76\footnotemark[3] & 326.26  & 58.8 & & 1.20 & 0.02 & & 25.0 & 0.20 &  2.90 & 2.50 & & 4.0 & 7.3 & 3900  \\
 63.50\footnotemark[2] & 334.00  & 31.0 & & 1.15 & & &  81.0 & 0.25 & 2.0 &  3.60 & & 2.5 & 5.0 & 3200  \\
 73.60\footnotemark[2] & 344.10  & 79.0 &  & 2.30 & & &  100.0 & 0.60 & 2.75 &  5.05 & 1.10 & 3.0 & 4.8 & 2600 \\
 79.73\footnotemark[3] & 350.23  & 79.9 &  & 4.90 & && 30.0 & 1.20 & 11.9 &  6.80 & & 4.5 & 7.1 & 2600  \\
 84.79\footnotemark[3] & 355.29  & 142.0 &  & 7.35  & 0.01 & 0.50 & 10.0 & 0.65 & 9.15 &  5.45 & & 3.5 & 6.2 & 2400 \\
 86.50\footnotemark[2] & 357.00  & 223.0 & & 3.15 &  & 0.10 & 515.0 & 0.70 & 7.90 &  13.1 & & 4.0 & 5.2 & 2250  \\
 98.60\footnotemark[2] & 369.10  & 72.25 && 6.05 & && 50.0 & 0.80 & 3.10 &  4.70 & 1.10 & 2.5 & 4.5 & 1900  \\
 103.70\footnotemark[3] & 374.20  & 219.0 && 32.05 & && 80.0 & 2.35 & 37.7 &  25.3 & 5.50 & 4.5 & 5.5 & 1800  \\
\hline
\end{tabular}
 \begin{tablenotes}
	\item{Source of spectra:}
       \item[a]{$^a$ This paper.}
       \item[b]{$^b$ \citet{sahu04et}}
       \item[c]{$^c$ \citet{maguire04et}}
     \end{tablenotes}
\end{minipage} 
\end{table*}

\begin{table*}
\centering
 \begin{minipage}{168mm}
\contcaption{ for SN 2005cs. (The second column is JD-2,453,000).}
  \smallskip 
  \addtocounter{footnote}{1}
  \begin{tabular}{@{}ccccccccccccccccc@{}}
\hline
 \multicolumn{17}{c}{SN 2005cs}\\
\hline 
phase & JD &  \multicolumn{12}{c}{$\tau_{ref}$} & n & $T_{bb}$ & $v_{phot}$  \\
  (days) & (days)  & \mbox{H\,{\sc i}} & \mbox{He\,{\sc i}} &  \mbox{N\,{\sc ii}} &  \mbox{Na\,{\sc i}} & \mbox{Mg\,{\sc i}}  & \mbox{Si\,{\sc i}} & \mbox{Si\,{\sc ii}} & \mbox{Ca\,{\sc ii}} & \mbox{Sc\,{\sc ii}} & \mbox{Ti\,{\sc ii}}  & \mbox{Fe\,{\sc ii}} & \mbox{Ba\,{\sc ii}} & & (kK) & (km s$^{-1})$  \\
\hline
 3.44 & 552.44  & 3.90   & 0.30 & 0.10 &&&&&&&&&& 4.0 & 15.0 & 7100  \\
 4.41 & 553.41  & 7.00   & 0.40 & 0.10 & &&& 0.10 &&&&&& 4.5 & 15.0 & 6900  \\
 5.39 & 554.39  & 10.6  & 0.30 & 0.10 &&&&&&&&&& 5.5 & 13.6 & 6500  \\
 8.43 & 557.43  & 15.3  & 0.20 &  &&&& 0.50 && & & & 0.35 &  5.0 & 10.6 &  5900  \\
 8.84 & 557.84  & 9.00   & 0.30 &  &&&& 0.60 & && & & 0.45 &  9.5 & 9.8 & 5950  \\
 14.36 & 563.36 & 81.0   & 1.50 & & 0.40 &&& 0.60 & 30.0 & 0.10 & 0.55 &  4.25 & 0.30 & 8.0 & 10.2& 4150  \\
 14.44 & 563.44  & 49.0  & 1.00  &  & 0.25 & 0.35 & & 0.50 & 27.0 & 0.20 & 0.90 &  1.30 & 0.35 & 7.0 & 9.8 & 4600  \\
 17.35 & 566.35  &  124.0 &  & & 0.60 & 0.70 & & 0.40 & 80.0 & 0.05 & 1.70 &  2.05 & & 8.0 & 8.4 & 4150  \\
 18.42 & 567.42  & 168.0 & & & 0.90 & 0.50 &&  0.70 & 5.00 & 0.50 & 1.40 &  1.50 & & 8.5 & 9.6 & 4150  \\
 22.45 & 571.45  & 36.0 &  &  & 0.30 & 1.00 & & 0.40 & & 0.70 & 1.20 &  1.05 & 0.90& 8.0 & 7.5 & 3550  \\
 34.44 & 583.44  & 52.0  &  &   & 5.10 & 0.40 & & 0.30 & 800.0 & 2.30 & 6.90 &  9.0 & 4.0 &5.0 & 6.5 & 1900  \\
 36.40 & 585.40  & 35.0  &  && 4.75 & 0.80  & & 0.30 & 100.0 & 1.85 & 8.85 &   10.3 & 1.95 & 5.0 & 6.2 & 1800 \\
 44.40 & 593.40  & 23.0  &  && 23.9 &  & 0.01 & 0.35 & 200.0 & 4.75 & 25.45 &  24.1 & 7.85 & 5.0 & 6.2 & 1350  \\
 61.40 & 610.40  & 15.0  &  && 119.9 &  & 0.04 & 0.25 & 800.0 & 12.35 & 73.50 &  54.4 & 63.40 & 5.5 & 5.7 & 1050 \\
\hline
\end{tabular}
 \begin{tablenotes}
	\item{}
       \item[a]{}
       \item[b]{}
       \item[c]{}
     \end{tablenotes}
\end{minipage} 
\end{table*}

\begin{table*}
\centering
 \begin{minipage}{168mm}
\contcaption{ for SN 2006bp}
  \smallskip 
  \addtocounter{footnote}{1}
  \begin{tabular}{@{}ccccccccccccccc@{}}
\hline
 \multicolumn{15}{c}{SN 2006bp} \\
\hline
phase & JD-2,451,000 &  \multicolumn{10}{c}{$\tau_{ref}$} & n & $T_{bb}$ & $v_{phot}$  \\
  (days) & (days)  & \mbox{H\,{\sc i}} & \mbox{He\,{\sc i}} &  \mbox{N\,{\sc ii}} &  \mbox{Na\,{\sc i}}  & \mbox{Si\,{\sc ii}} & \mbox{Ca\,{\sc ii}} & \mbox{Sc\,{\sc ii}} & \mbox{Ti\,{\sc ii}}  & \mbox{Fe\,{\sc ii}} & \mbox{Ba\,{\sc ii}} & & (kK) & (km s$^{-1}$)  \\
\hline 
 5.30 & 840.30  & 2.15 & 0.15 &&&&&&&&& 4.0 & 10.2 & 12050  \\
 7.10 & 842.10  & 3.20 & 0.15 &&&&&&&&& 3.5 & 12.0 & 10750  \\
 9.10 & 844.10  & 6.05 & 0.15 &&&&&&&&& 4.5 & 11.7 & 9600  \\
 11.11 & 846.11  & 10.35 & 0.20 &&&&&&&&& 5.5 & 12.7  & 9550  \\
 15.10 & 850.10  & 15.65 & 0.05 & 0.04  & 0.20 &   0.80 &  & && 0.35 & & 4.5 & 10.8 & 8750  \\
 20.28 & 855.28  & 26.70 & & 0.03 &  0.20 &   0.40 & 23.0 & 0.08 & 0.20 &   1.30 & & 4.5 & 9.5 & 8100  \\
 24.26 & 859.26  & 20.00 & 0.10 & 0.01  & 0.10 &   0.10 & 25.0 & 0.10 & 0.80 &   1.50 & 0.50 & 5.5 & 8.5 & 7050  \\
 32.26 & 867.26  & 59.70 & 1.50 & & 0.55 &  &  129.0 &  0.30 & 1.65 &  1.75 & 0.20 & 6.0 & 8.0 & 6150 \\
 41.21 & 876.21  & 118.6 & 2.00 & &  1.25 & &   326.0 & 0.65 & 3.50 &  2.60 & 0.85 & 7.0 & 7.2 & 5050  \\
 56.19 & 891.19  & 145.1 & 1.00 &   4.65 &  & & 300.0 & 0.20 & 6.40 &  5.35 & 0.20 & 6.0 & 7.2 & 4050 \\
 72.04 & 907.14  & 150.1 & &  & 18.4 &  & 165.0 & 1.90 & 10.85 & 6.75 & 1.45 & 5.5 & 7.0 & 3850  \\
\hline
\end{tabular}
 \begin{tablenotes}
	\item{}
       \item[a]{}
       \item[b]{}
       \item[c]{}
     \end{tablenotes}
\end{minipage}
\end{table*}

\label{lastpage}
\end{document}